\documentclass[runningheads]{llncs}
\usepackage[T1]{fontenc}
\usepackage{graphicx}
\usepackage[colorlinks,urlcolor=blue]{hyperref}  %
\usepackage{color}
\usepackage{url}

\usepackage{algorithm,algpseudocode}

\usepackage{mathtools}
\usepackage{amsmath,amsfonts,amssymb}
\usepackage{tabularray}
\usepackage{booktabs}
\usepackage{upgreek}
\usepackage{enumitem}
\usepackage{bm}
\usepackage{censor}
\usepackage{textcomp}
\usepackage[english]{babel}
\usepackage{lastpage}
\usepackage[section]{placeins}
\usepackage{fancyvrb} 
\usepackage[dvipsnames,table]{xcolor}
\usepackage{fontawesome5}

\usepackage{multirow}
\usepackage{makecell}
\usepackage{subfigure}

\newcommand*{\Scale}[2][4]{\scalebox{#1}{$#2$}}%

\definecolor{LightCyan}{rgb}{0.88,1,1}
\definecolor{lightskyblue}{RGB}{225, 235, 240}

\definecolor{Gray}{gray}{0.90}
\definecolor{white}{rgb}{1.0, 1.0, 1.0}

\definecolor{Lightgreen}{RGB}{218, 246, 230 }

\usepackage{graphics,color}
\usepackage{xcolor}
\definecolor{label1}{rgb}{0.76,0.59,0.77}
\definecolor{label2}{rgb}{0.28,0.5,0.72}
\definecolor{label3}{rgb}{0.33,0.63,0.36}
\definecolor{label4}{rgb}{0.79,0.4,0.17}
\definecolor{label5}{rgb}{0.94,0.53,0.2}
\definecolor{label6}{rgb}{0.72,0.86,0.59}
\definecolor{label7}{rgb}{1,1,0.65}
\definecolor{label8}{rgb}{0.93,0.62,0.61}
\definecolor{label9}{rgb}{0.4,0.15,0.33}
\definecolor{label10}{rgb}{0.75,0.21,0.29}
\definecolor{label11}{rgb}{0.35,0.73,0.8}
\definecolor{label12}{rgb}{0.94,0.9,0.32}
\definecolor{label13}{rgb}{0.96,0.76,0.48}

\newsavebox{\spleen}
\savebox{\spleen}{\textcolor{label1}{\rule{1.5in}{1.5in}}}

\newsavebox{\rkid}
\savebox{\rkid}{\textcolor{label2}{\rule{1.5in}{1.5in}}}

\newsavebox{\lkid}
\savebox{\lkid}{\textcolor{label3}{\rule{1.5in}{1.5in}}}

\newsavebox{\gall}
\savebox{\gall}{\textcolor{label4}{\rule{1.5in}{1.5in}}}

\newsavebox{\eso}
\savebox{\eso}{\textcolor{label5}{\rule{1.5in}{1.5in}}}

\newsavebox{\liver}
\savebox{\liver}{\textcolor{label6}{\rule{1.5in}{1.5in}}}

\newsavebox{\sto}
\savebox{\sto}{\textcolor{label7}{\rule{1.5in}{1.5in}}}

\newsavebox{\aorta}
\savebox{\aorta}{\textcolor{label8}{\rule{1.5in}{1.5in}}}

\newsavebox{\ivc}
\savebox{\ivc}{\textcolor{label9}{\rule{1.5in}{1.5in}}}

\newsavebox{\veins}
\savebox{\veins}{\textcolor{label10}{\rule{1.5in}{1.5in}}}

\newsavebox{\panc}
\savebox{\panc}{\textcolor{label11}{\rule{1.5in}{1.5in}}}

\newsavebox{\rad}
\savebox{\rad}{\textcolor{label12}{\rule{1.5in}{1.5in}}}

\newsavebox{\lad}
\savebox{\lad}{\textcolor{label13}{\rule{1.5in}{1.5in}}}

\begin{document}

\title{Frequency Domain Adversarial Training for Robust Volumetric Medical Segmentation}
\titlerunning{Volumetric Adversarial Frequency Attack and Training (VAFA \& VAFT)}

\author{Asif Hanif\inst{1} \and Muzammal Naseer\inst{1} \and Salman Khan\inst{1}  \and Mubarak Shah\inst{2} \and \\Fahad Shahbaz Khan\inst{1,3}
} %

\authorrunning{A. Hanif et al.}

\institute{Mohamed Bin Zayed University of Artificial Intelligence (MBZUAI), UAE
\email{\{asif.hanif,muzammal.naseer,salman.khan,fahad.khan\}@mbzuai.ac.ae}
\and
University of Central Florida (UCF), USA\\
\email{shah@crcv.ucf.edu} \\
\and
Link\"{o}ping University, Sweden
}

\maketitle              %

\begin{abstract}
It is imperative to ensure the robustness of deep learning models in critical applications such as, healthcare. While recent advances in deep learning have improved the performance of volumetric medical image segmentation models, these models cannot be deployed for real-world applications immediately due to their vulnerability to adversarial attacks. We present a 3D frequency domain adversarial attack for volumetric medical image segmentation models and demonstrate its advantages over conventional input or voxel domain attacks. Using our proposed attack, we introduce a novel frequency domain adversarial training approach for optimizing a robust model against voxel and frequency domain attacks.  Moreover, we propose frequency consistency loss to regulate our frequency domain adversarial training that achieves a better tradeoff between model's performance on clean and adversarial samples. Code is available at \url{https://github.com/asif-hanif/vafa}.
\end{abstract}

\section{Introduction} \label{s:intro}

Semantic segmentation of organs, anatomical structures, or anomalies in medical images (e.g. CT or MRI scans) remains one of the fundamental tasks in medical image analysis. %
Volumetric medical image segmentation (MIS) helps healthcare professionals to diagnose conditions more accurately, plan medical treatments, and perform image-guided procedures. %
Although deep neural networks (DNNs) have shown remarkable improvements in performance for different vision tasks, including volumetric MIS, their real-world deployment is not straightforward particularly due to the  vulnerabilities towards adversarial attacks \cite{szegedy2013intriguing}. An adversary can deliberately manipulate input data by crafting and adding perturbations to the input that are imperceptible to the human eye but cause the DNN to produce incorrect outputs \cite{goodfellow2014explaining}. %
Adversarial attacks pose a serious security threat to DNNs \cite{akhtar2018threat}, as they can be used to cause DNNs to make incorrect predictions in a wide range of applications, including %
DNN-based medical imaging systems. To mitigate these threats, various techniques have been explored, including adversarial training, input data transformations, randomization, de-noising auto-encoders, feature squeezing, and robust architectural changes \cite{akhtar2018threat}. Although significant progress has been made in adversarial defenses, however, this area is still evolving due to the development of attacks over time \cite{carlini2017adversarial}. 

Ensuring the adversarial robustness of the models involved in safety-critical applications such as, medical imaging and healthcare is of paramount importance because a misdiagnosis or incorrect decision can result in life-threatening implications. Moreover, the weak robustness of deep learning-based medical imaging models will create a trust deficit among clinicians, making them reluctant to rely on the model predictions. %
The adversarial robustness of the medical imaging models is still an open and under-explored area \cite{ma2021understanding,daza2021towards}. Furthermore, most adversarial attacks and defenses have been designed for 2D natural images and little effort has been made to secure  volumetric (3D) medical data \cite{ma2021understanding}. %

\begin{figure}[!t]
\includegraphics[trim={0 2cm 0 0}, width=\textwidth]{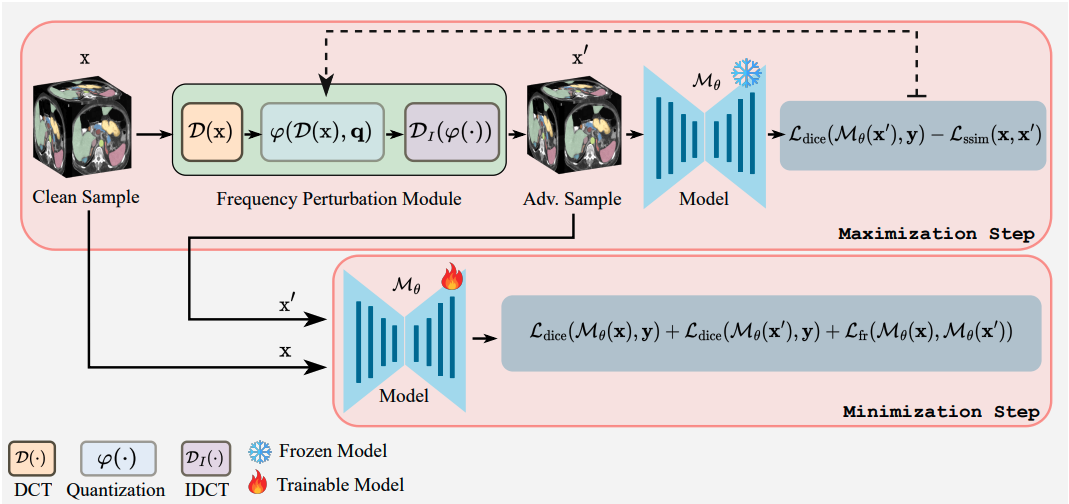}
\caption{\small \textbf{Overview of Adversarial Frequency Attack and Training}: A model trained on voxel-domain adversarial attacks is vulnerable to frequency-domain adversarial attacks. In our proposed adversarial training method, we generate adversarial samples by perturbing their frequency-domain representation using a novel module named "Frequency Perturbation".  The model is then updated while minimizing the dice loss on clean and adversarially perturbed images. Furthermore, we propose a frequency consistency loss to improve the model performance.} \label{fig: main-figure}
\end{figure}

In the context of 2D natural images, it has been recently observed that frequency-domain based adversarial attacks are more effective against the defenses that are primarily designed to ``undo'' the impact of pixel-domain adversarial noise in natural images \cite{duan2021advdrop}. Motivated by this observation in 2D natural images, here we explore the effectiveness of frequency-domain based adversarial attacks in the regime of volumetric medical image segmentation and aim to obtain a volumetric MIS model that is robust against %
adversarial attacks. To achieve this goal, we propose a \textit{min-max} objective for adversarial training of \textit{volumetric MIS model in frequency-domain}. For \textit{maximization} step, we introduce \textbf{V}olumetric \textbf{A}dversarial \textbf{F}requency \textbf{A}ttack - \textbf{VAFA} (Fig. \ref{fig: main-figure}, Sec. \ref{subsec: Adversarial Frequency Attack}) which operates in the frequency-domain of the data (unlike other prevalent voxel-domain attacks) and explicitly takes into account the 3D nature of the volumetric medical data to achieve higher fooling rate. %
For \textit{minimization} step, we propose \textbf{V}olumetric \textbf{A}dversarial \textbf{F}requency-domain  \textbf{T}raining - \textbf{VAFT} (Fig. \ref{fig: main-figure}, Sec. \ref{subsec: Adversarial Frequency Training}) to obtain a model that is robust to %
adversarial attacks. In VAFT, we update model parameters on clean and adversarial (obtained via VAFA) samples and further introduce a novel \textit{frequency consistency loss} to keep frequency representation of the logits of clean and adversarial samples close to each other for a better accuracy tradeoff. In summary, our contributions are as follows:
\begin{itemize}
     \item We propose an approach with a min-max objective for adversarial training of volumetric MIS model in the frequency domain. In the maximization step, we introduce a volumetric adversarial frequency attack (VAFA) that is specifically designed for volumetric medical data to achieve higher fooling rate. Further, we introduce a volumetric adversarial frequency-domain training (VAFT) based on a frequency consistency loss in the minimization step to produce a model that is robust to %
     adversarial attacks.  
     \item We conduct experiments with two different hybrid CNN-transformers based volumetric medical segmentation methods for multi-organ segmentation.
 \end{itemize}
\noindent \textbf{Related Work:} There are three main types of popular volumetric MIS model architectures: CNN \cite{ronneberger2015u}, Transformer \cite{karimi2021convolution} and hybrid \cite{hatamizadeh2022unetr,shaker2022unetr++}. Research has shown that medical machine learning models can be manipulated in various ways by an attacker, such as adding imperceptible perturbation to the image, rotating the image, or modifying medical text \cite{finlayson2019adversarial}. Adversarial attack studies on medical data have primarily focused on classification problems and voxel-domain adversaries. %
For example, Ma \textit{et al.} \cite{ma2021understanding} have used four types of pixel-domain attacks \cite{madry2017towards,goodfellow2014explaining,kurakin2018adversarial,carlini2017towards} on two-class and multi-class medical datasets.
Li \textit{ et al.} \cite{li2019volumetric}  and Daza \textit{et al.} \cite{daza2021towards} have focused on single-step and iterative  adversarial attacks  \cite{goodfellow2014explaining,kurakin2016adversarial,croce2020reliable} on the volumetric MIS. In constant to voxel-domain adversarial attacks, our approach works in the frequency-domain. %

\section{Frequency Domain Adversarial Attack and Training}

We aim to train a model for volumetric medical segmentation that is robust against adversarial attacks. Existing adversarial training (AT) approaches rely on min-max optimization \cite{madry2017towards,goodfellow2014explaining,kurakin2018adversarial} and operate in the input space. They find adversaries by adding the adversarial perturbation to the input samples by maximizing the model loss (e.g., dice loss in segmentation). The loss function is then minimized on such adversaries to update the model parameters. %
In this work, we propose a frequency-domain adversarial attack that takes into account the 3D nature of the volumetric medical data and performs significantly better than the other voxel-domain as well as 2D frequency domain attacks (Tab.~\ref{tbl:afa3d-afa2d-voxel-attacks}). Based on our attack, we then introduce a novel frequency-domain adversarial training to make the model resilient to adversarial attacks. Additionally, we observe that our approach improves/retains the performance of the robust model on clean samples when compared to the non-robust model. Our approach optimizes adversarial samples by perturbing the 3D-DCT coefficients within the frequency domain using our frequency perturbation module (Fig. \ref{fig: main-figure}) and adversarial guidance from the segmentation loss (Sec.~\ref{subsec: Adversarial Frequency Attack}). We find adversarial samples with high perceptual quality by maximizing the structural similarity between clean and adversarial samples. Using clean and adversarial samples, we propose updating the model parameters by simultaneously minimizing the segmentation loss (i.e. Dice loss) and the frequency consistency loss (Eq.~\ref{eq: fr loss}) between the clean and adversarial outputs of the segmentation model. %

\noindent \textbf{3D Medical Segmentation Framework}: 
Deep learning-based 3D medical segmentation generally uses encoder-decoder architectures \cite{lei2020medical}. %
The encoder produces a latent representation of the input sample.
A segmentation map of the input sample is generated by the decoder using the latent feature representation. The decoder usually incorporates skip connections from the encoder to preserve spatial information \cite{hatamizadeh2022unetr}. %
Next, we describe our proposed volumetric frequency-domain adversarial attack in Sec.~\ref{subsec: Adversarial Frequency Attack} and then training in Sec.~\ref{subsec: Adversarial Frequency Training}.%

\subsection{Volumetric Adversarial Frequency Attack (VAFA)}
\label{subsec: Adversarial Frequency Attack}

\begin{figure*}[!t]

\begin{minipage}{\textwidth}
\centering
\begin{minipage}{\linewidth}
\scalebox{0.06}{{\usebox{\spleen}}} \small Spleen ~~~\scalebox{0.06}{{\usebox{\rkid}}} \small R-Kidney ~~~ \scalebox{0.06}{{\usebox{\lkid}}} \small L-Kidney ~~~\scalebox{0.06}{{\usebox{\gall}}} \small Gallbladder ~~~\scalebox{0.06}{{\usebox{\eso}}} \small Esophagus ~~~\scalebox{0.06}{{\usebox{\liver}}} \small Liver ~~~ \scalebox{0.06}{{\usebox{\sto}}} \small Stomach ~\scalebox{0.06}{{\usebox{\aorta}}} \small Aorta ~~~~~~~~~ \scalebox{0.06}{{\usebox{\ivc}}} \small IVC ~~~~~~~~~\scalebox{0.06}{{\usebox{\veins}}} \small Veins ~\scalebox{0.06}{{\usebox{\panc}}} \small Pancreas ~~~ \scalebox{0.06}{{\usebox{\rad}}} \small Rad. ~~~ \scalebox{0.06}{{\usebox{\lad}}} \small Lad.
\end{minipage}
\\
\vspace{0.2em}
\begin{minipage}{0.16\textwidth}
      \centering
    \includegraphics[height=2.2cm, width=\linewidth, trim={1cm 8cm 0.5cm 1cm},clip ]{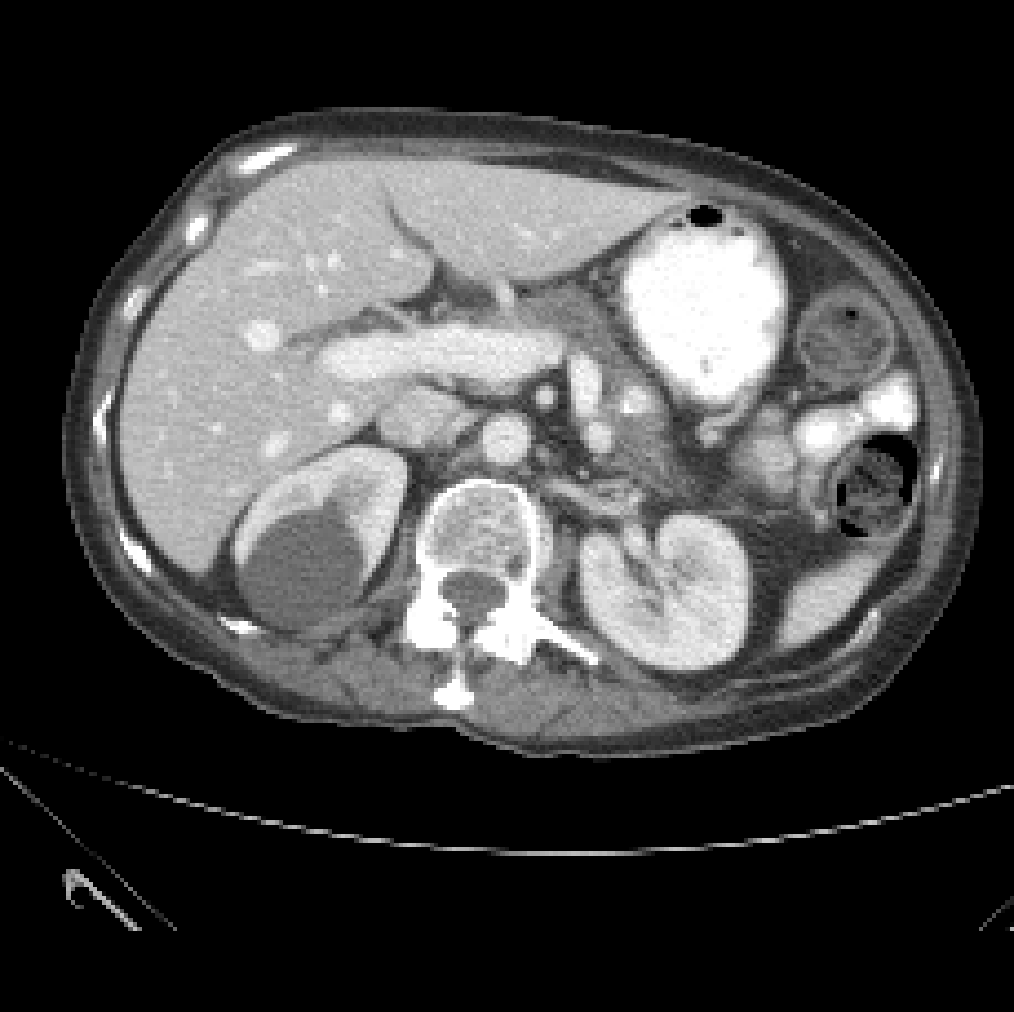}
  \end{minipage}
  \begin{minipage}{0.16\textwidth}
      \centering
   \includegraphics[height=2.2cm, width=\linewidth,  trim={1cm 8cm 0.5cm 1cm},clip ]{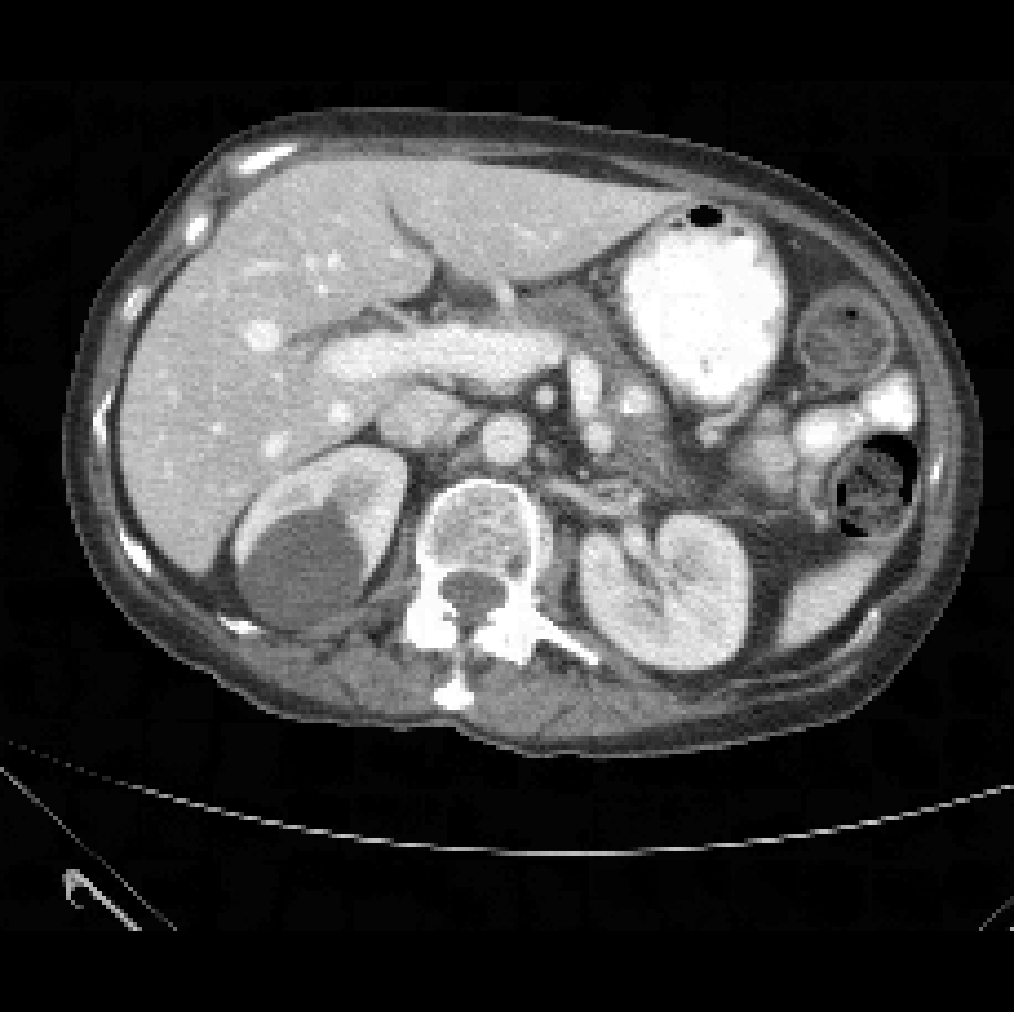}
  \end{minipage}
\begin{minipage}{0.16\textwidth}
      \centering
  \includegraphics[height=2.2cm, width=\linewidth,  trim={1cm 8cm 0.5cm 1cm},clip ]{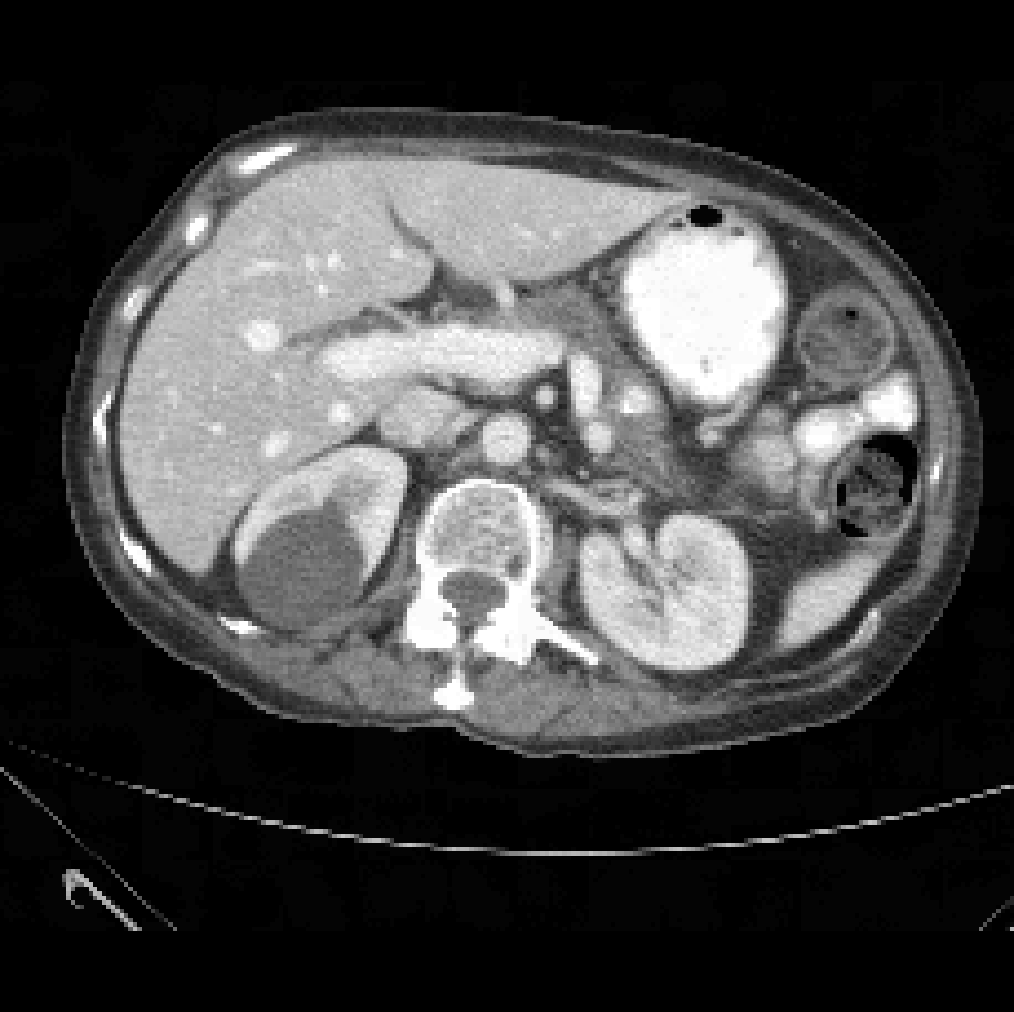}
  \end{minipage}
  \begin{minipage}{0.16\textwidth}
      \centering
    \includegraphics[height=2.2cm, width=\linewidth,  trim={1cm 8cm 0.5cm 1cm},clip ]{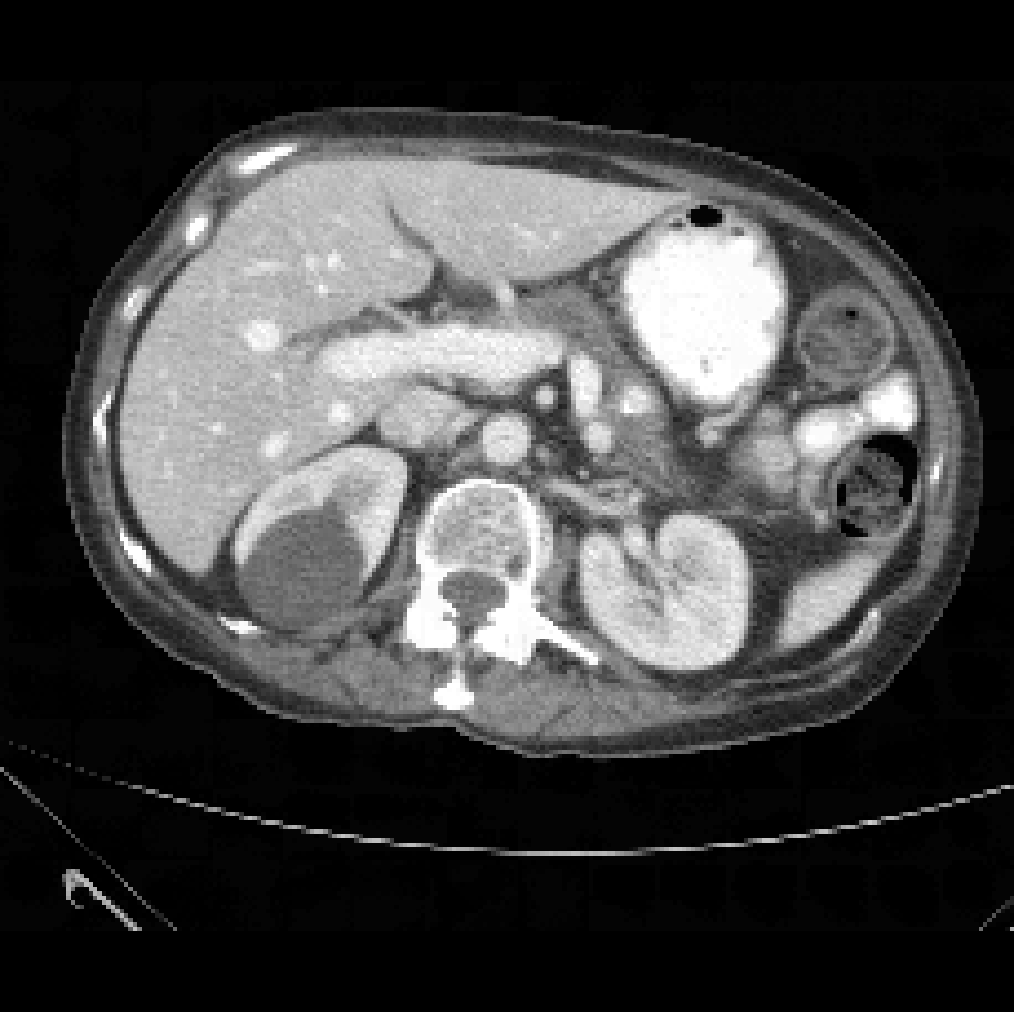}
  \end{minipage}
    \begin{minipage}{0.16\textwidth}
      \centering
   \includegraphics[height=2.2cm, width=\linewidth,  trim={1cm 8cm 0.5cm 1cm},clip ]{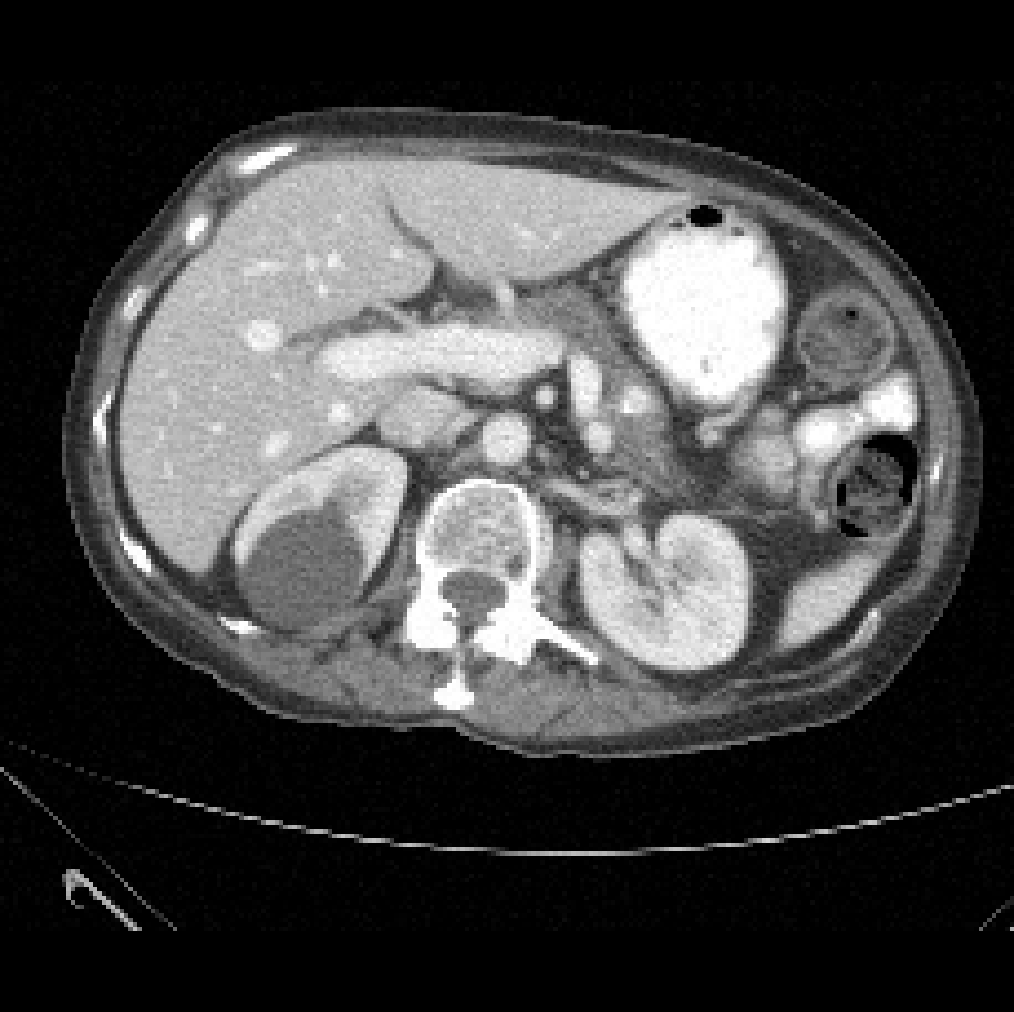}
  \end{minipage}
\begin{minipage}{0.16\textwidth}
      \centering
  \includegraphics[height=2.2cm, width=\linewidth,  trim={1cm 8cm 0.5cm 1cm},clip ]{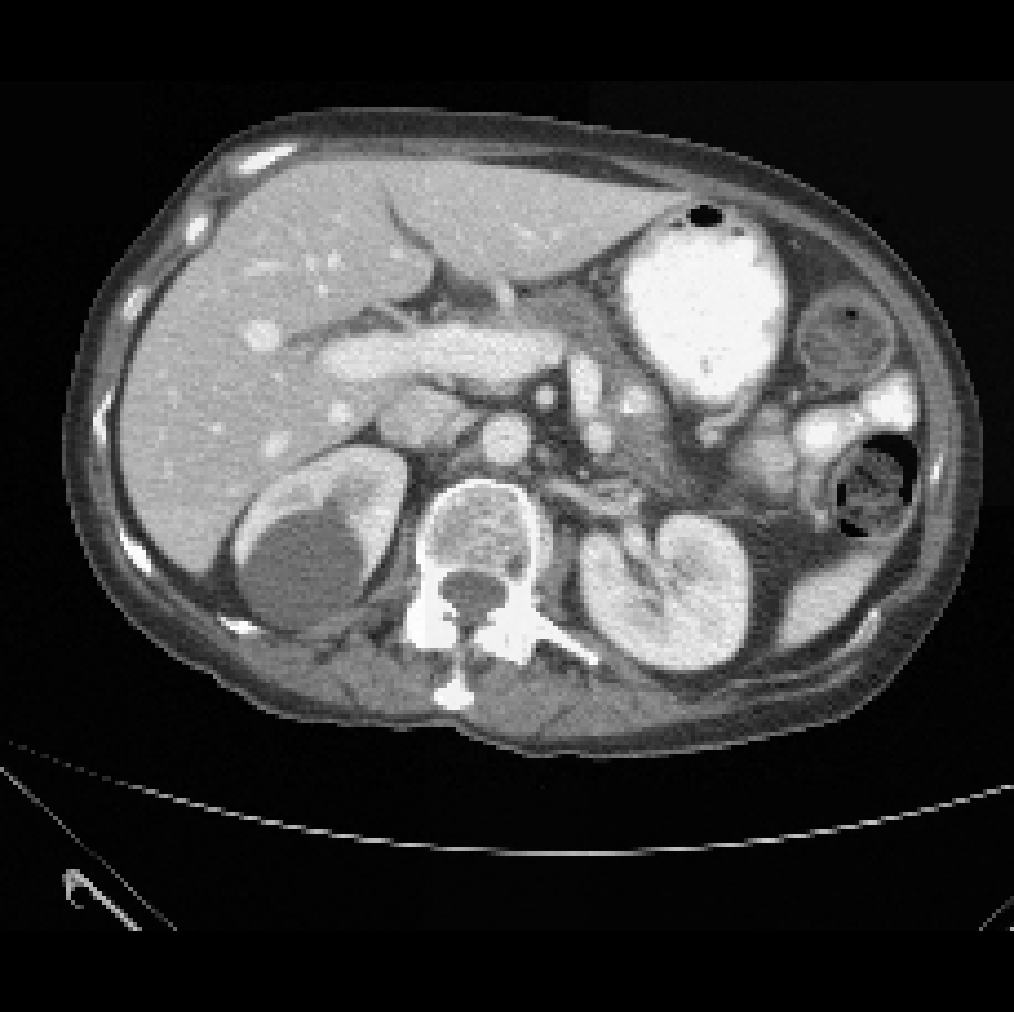}
  \end{minipage}
  \\
  \begin{minipage}{0.16\textwidth}
      \centering
   \includegraphics[height=2.2cm, width=\linewidth,,  trim={1cm 8cm 0.5cm 1cm},clip ]{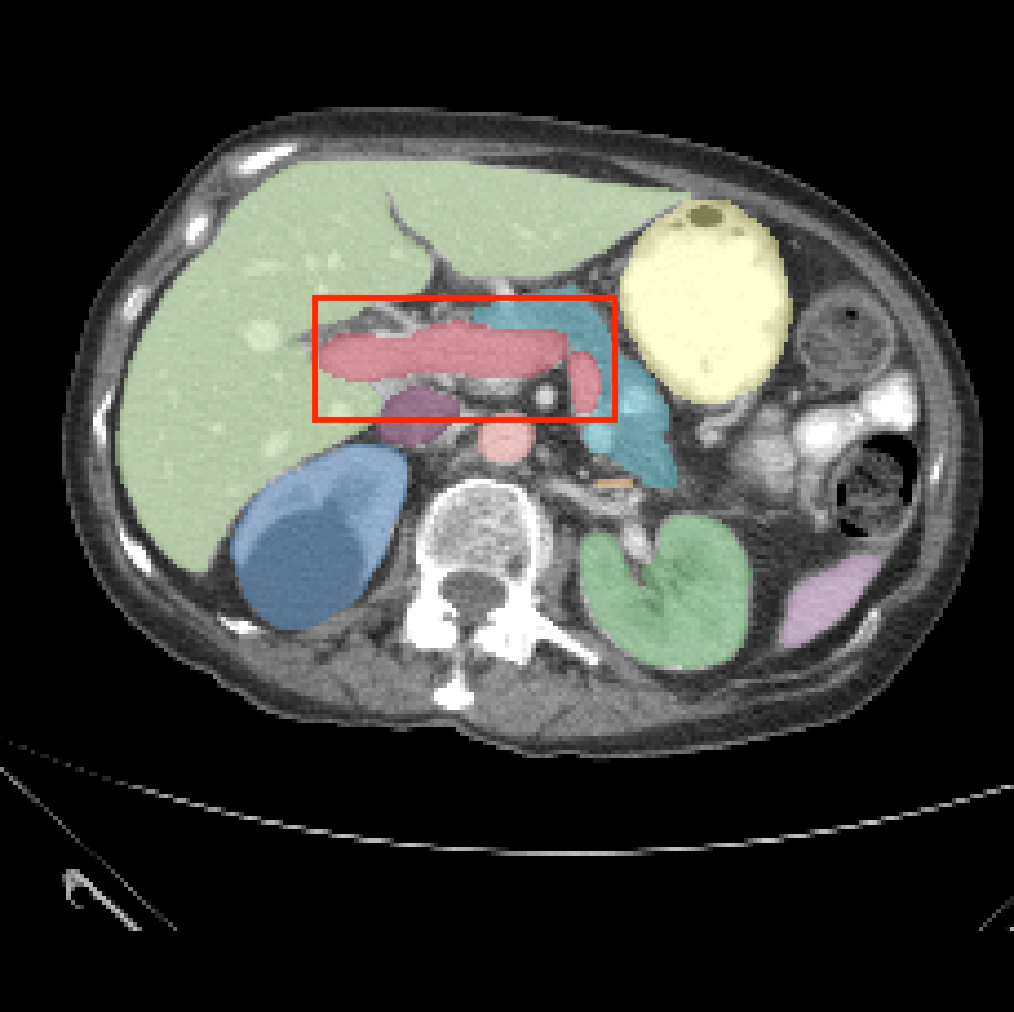}
   \footnotesize \textbf{Clean}
  \end{minipage}
\begin{minipage}{0.16\textwidth}
      \centering
  \includegraphics[height=2.2cm, width=\linewidth,  trim={1cm 8cm 0.5cm 1cm},clip ]{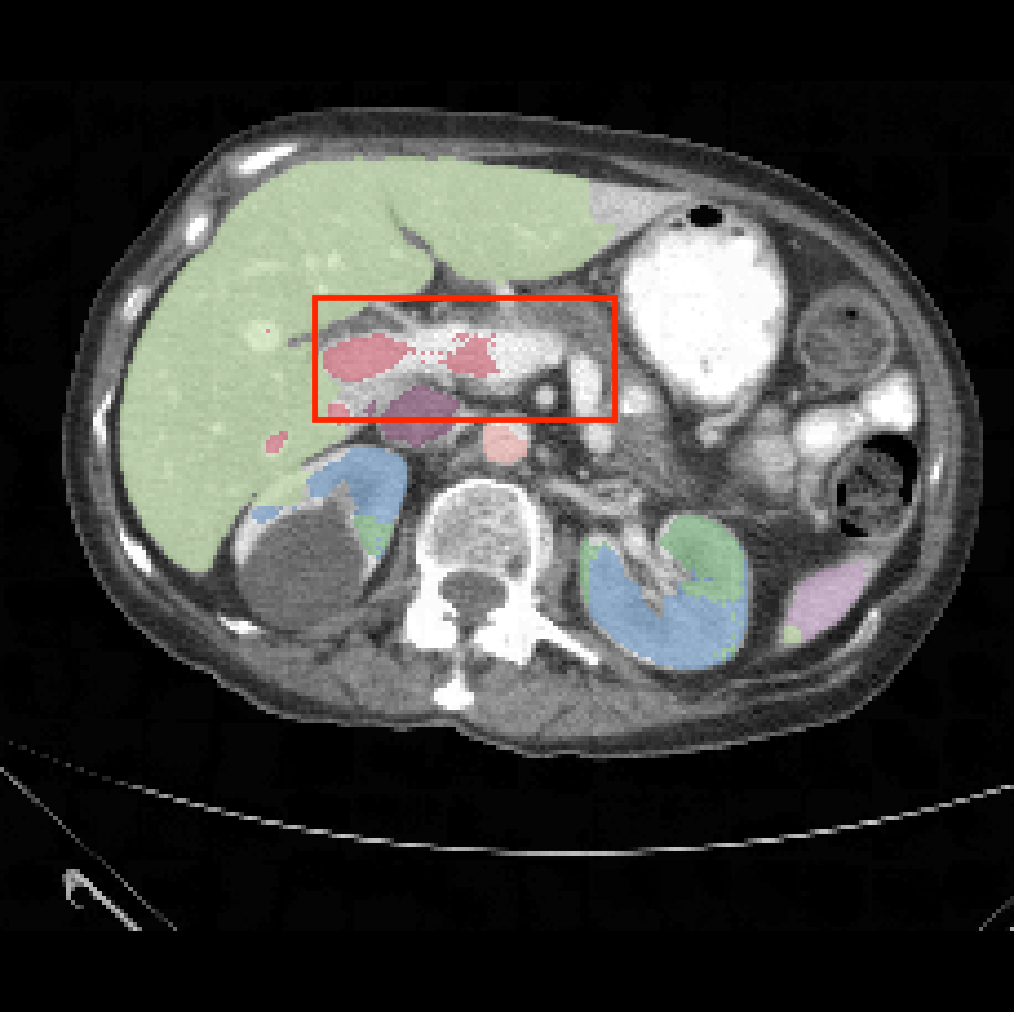}
  \footnotesize \textbf{PGD}
  \end{minipage}
  \begin{minipage}{0.16\textwidth}
      \centering
    \includegraphics[height=2.2cm, width=\linewidth,  trim={1cm 8cm 0.5cm 1cm},clip ]{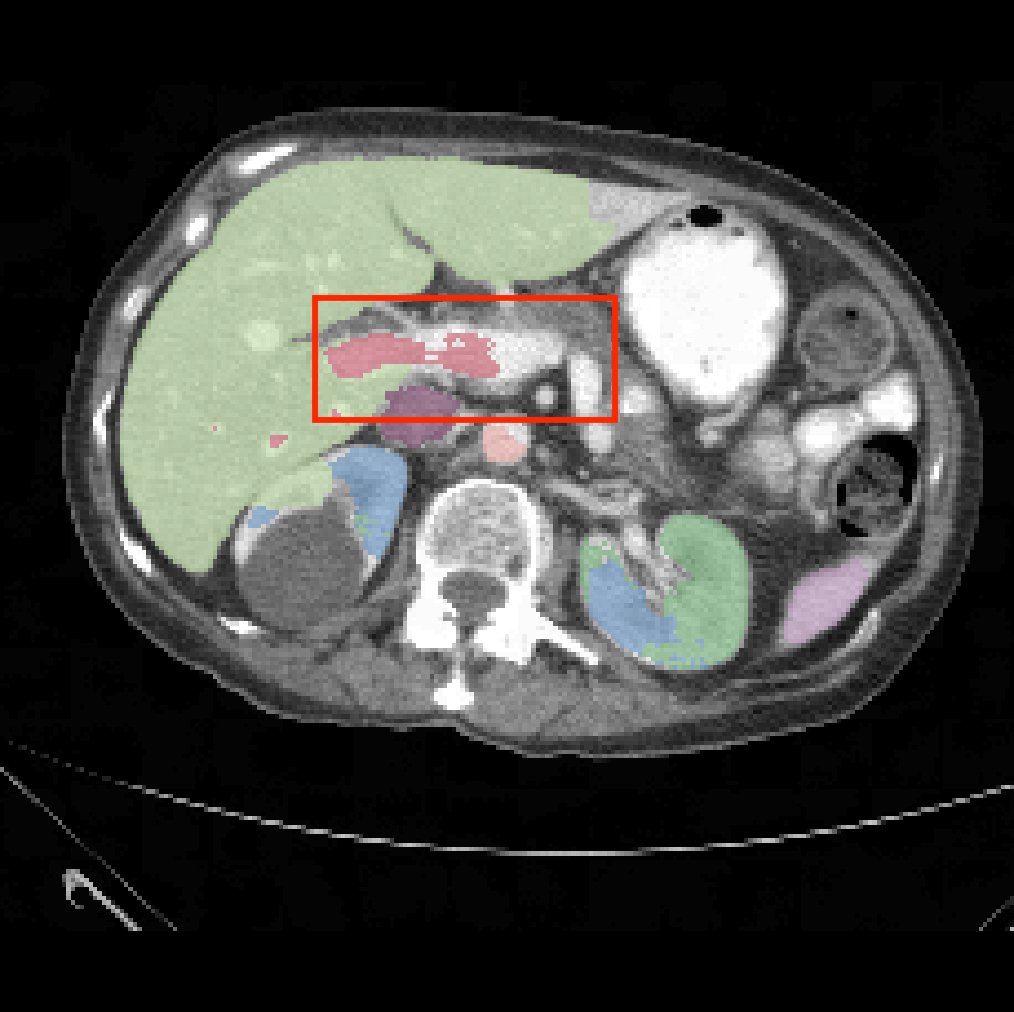}
    \footnotesize \textbf{FGSM}
  \end{minipage}
  \begin{minipage}{0.16\textwidth}
      \centering
   \includegraphics[height=2.2cm, width=\linewidth,  trim={1cm 8cm 0.5cm 1cm},clip ]{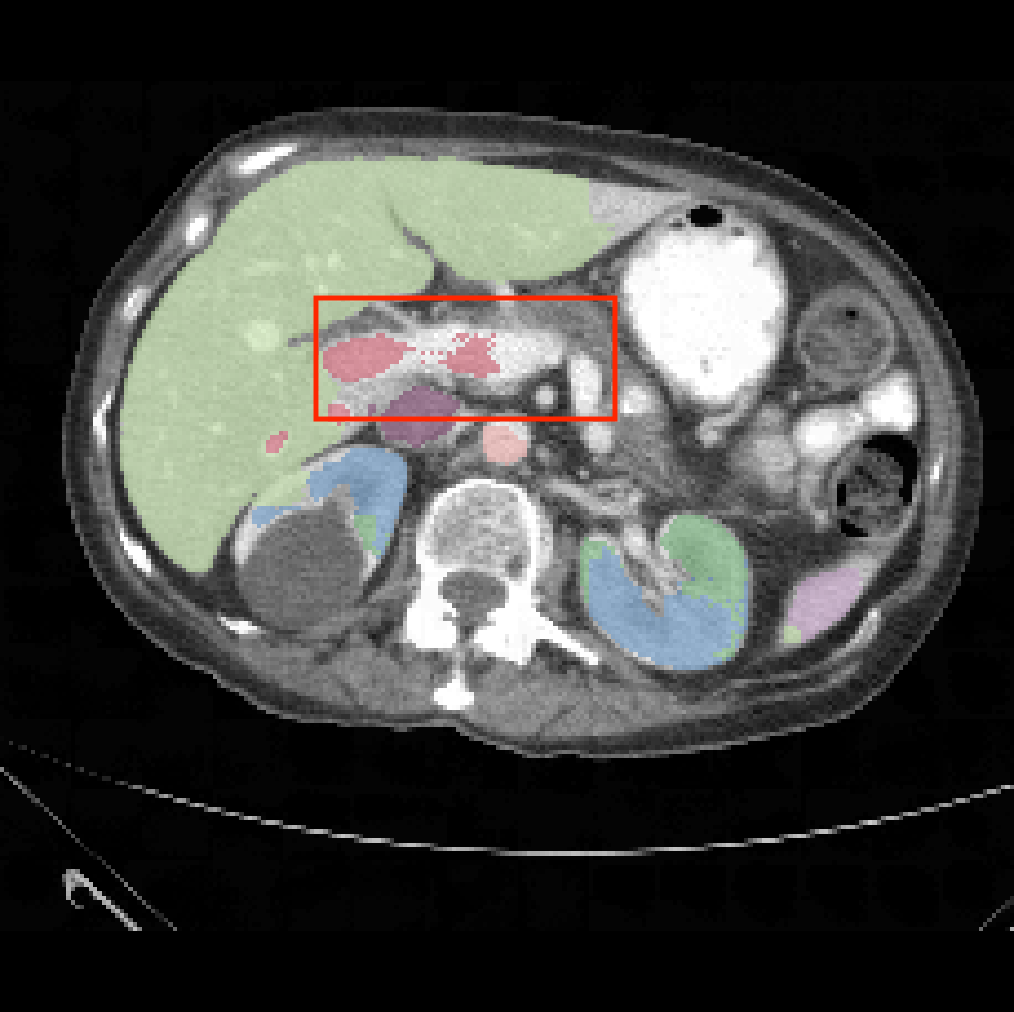}
   \footnotesize \textbf{BIM}
  \end{minipage}
\begin{minipage}{0.16\textwidth}
      \centering
  \includegraphics[height=2.2cm, width=\linewidth,  trim={1cm 8cm 0.5cm 1cm},clip ]{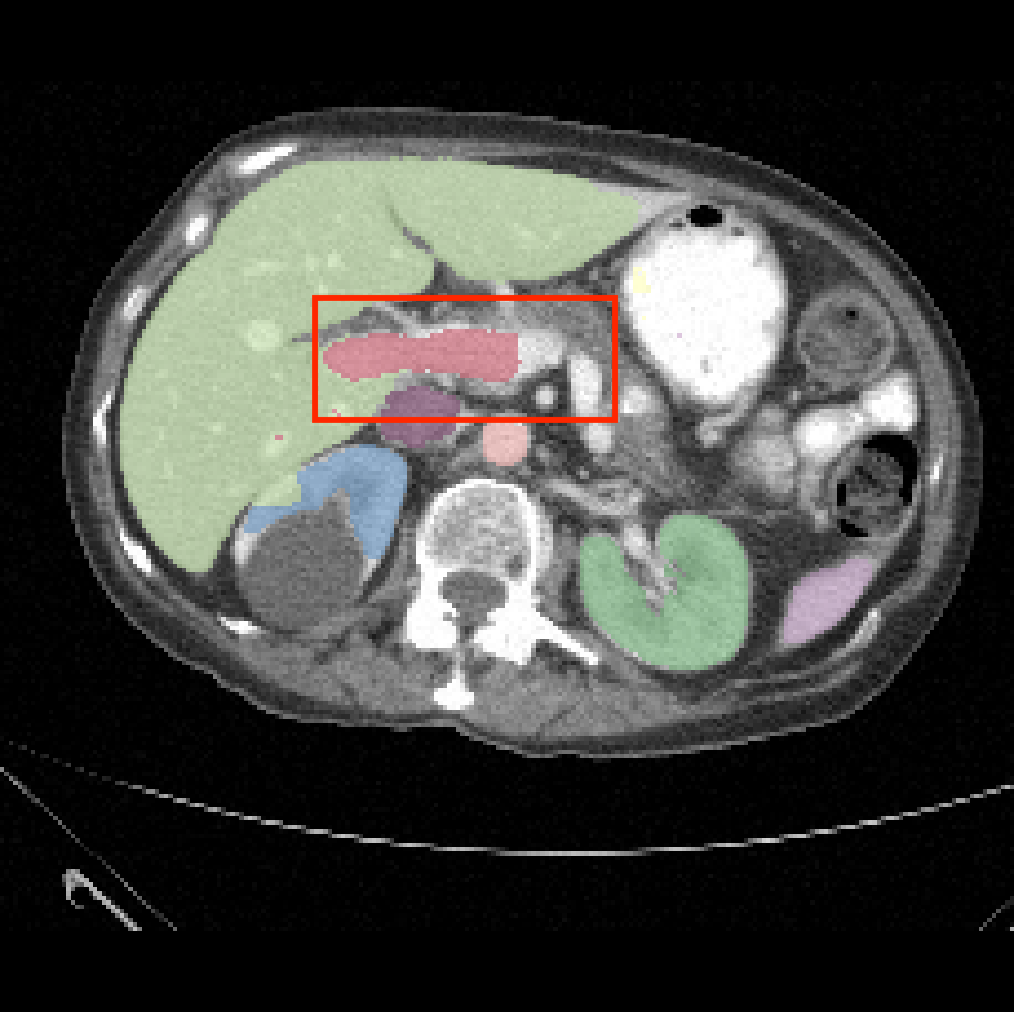}
  \footnotesize \textbf{GN}
  \end{minipage}
  \begin{minipage}{0.16\textwidth}
      \centering
    \includegraphics[height=2.2cm, width=\linewidth,  trim={1cm 8cm 0.5cm 1cm},clip ]{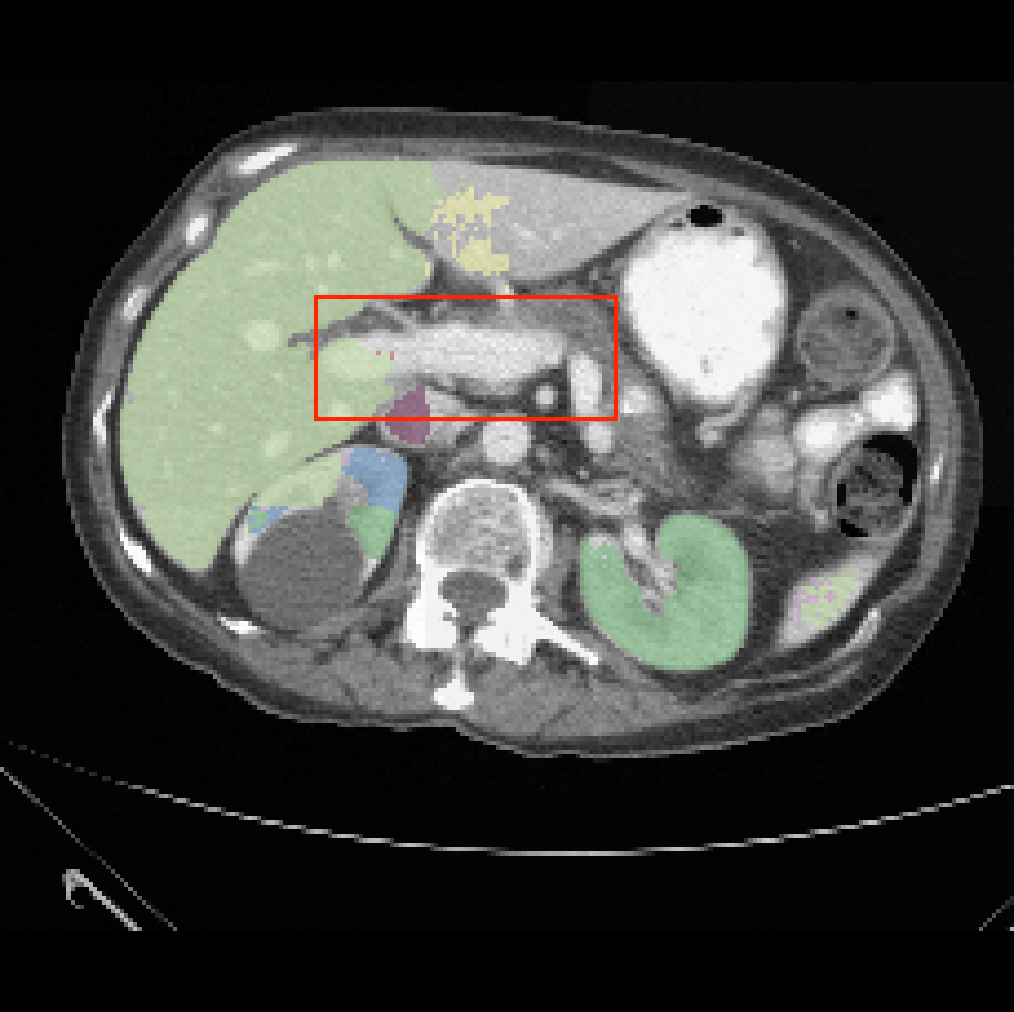}
    \footnotesize \textbf{VAFA{\tiny(ours)}}
  \end{minipage}
 
\caption{Qualitative multi-organ segmentation comparison under different attacks on the UNETR \cite{hatamizadeh2022unetr} model. Top row shows example images and bottom row shows the corresponding segmentation masks predicted by the model under different attacks. Compared to different voxel-domain attacks (PGD \cite{madry2017towards}, FGSM \cite{goodfellow2014explaining}, BIM \cite{kurakin2018adversarial} and GN \cite{kim2020torchattacks}), our attack (VAFA) achieves higher fooling rate (highlighted in red bounding box) while maintaining comparable perceptual similarity. Best viewed zoomed in.}
\label{fig: visualization}
\end{minipage}
\end{figure*}

Generally, adversarial attacks operate in the voxel domain by adding an imperceptible perturbation to the input data. In contrast, our attack perturbs the 3D-DCT coefficient to launch a frequency-domain attack for 3D medical image segmentation. %
Our \emph{Frequency Perturbation Module} (FPM) transforms voxel-domain data into frequency-domain by using discrete cosine transforms (DCTs) and perturbs the DCT coefficients using a \emph{learnable quantization}. It then takes an inverse DCT of the perturbed frequency-domain data and returns voxel-domain image. We keep the model in a ``frozen'' state while maximizing the dice loss \cite{sudre2017generalised} for segmentation and minimizing structural similarity loss \cite{wang2004image} for perceptual quality. 
We represent a 3D (volumetric) single channel clean sample by $\mathrm{X} \in \mathbb{R}^{1\times H\times W\times D}$ and its ground-truth binary segmentation mask by $\mathrm{Y} \in \{0,1\}^{\mathrm{NumClass} \times H\times W\times D}$, where $``\mathrm{NumClass}"$ is the number of classes. We split $\mathrm{X}$ into $n$ 3D patches  i.e. $\mathrm{X} \mapsto \{\bm{\mathrm{x}}_i\}_{i=1}^{n}$, where $\bm{\mathrm{x}}_i\in \mathbb{R}^{h\times w\times d}$ and $h\le H,w\le W, d\le D$, $h=w=d$. We apply our frequency perturbation module to each of these patches.

\noindent \textbf{Frequency Perturbation Module:} 
We apply a 3D discrete cosine transform (DCT), represented as $\mathcal{D}(\cdot)$, to each patch $\bm{\mathrm{x}}_i$. The resulting DCT coefficients are then processed through a function $\varphi(\cdot)$, which performs three operations: quantization, differentiable rounding (as described in \cite{gong2019differentiable}), and subsequent de-quantization. $\varphi(\cdot)$ utilizes a learnable quantization table $\bm{\mathrm{q}}\in\mathbb{Z}^{h\times w\times d}$ to modify the DCT coefficients, setting some of them to zero. In particular, $\textstyle \varphi(\mathcal{D}(\bm{\mathrm{x}}),\bm{\mathrm{q}}) \coloneqq \lfloor\frac{\mathcal{D}(\bm{\mathrm{x}})}{\bm{\mathrm{q}}}\rfloor \odot \bm{\mathrm{q}}$, where DCT coefficients of a patch (i.e. $\mathcal{D}(\bm{\mathrm{x}})$) are element-wise divided by quantization table $\bm{\mathrm{q}}$. After the division operation, the result undergoes rounding using a differentiable rounding operation \cite{gong2019differentiable}, resulting in some values being rounded down to zero. The de-quantization step involves element-wise multiplication of $\lfloor\frac{\mathcal{D}(\bm{\mathrm{x}})}{\bm{\mathrm{q}}}\rfloor$ with the same quantization table $\bm{\mathrm{q}}$. This step allows us to reconstruct the quantized DCT coefficients. Since quantization table is in the denominator of the division operation, therefore, higher quantization table values increase the possibility of more DCT coefficients being rounded down to zero. To control the number of DCT coefficients being set to zero, we can constrain the values of the quantization table to a maximum threshold (constraint in Eq. \ref{eq: afa objective}). In other words, $\varphi(\cdot)$ performs a 3D adversarial lossy compression on input through a learnable quantization table. Finally, a 3D inverse DCT (IDCT) is performed on the output of $\varphi(\cdot)$ in order to obtain an adversarially perturbed voxel-domain representation, denoted by $\bm{\mathrm{x}}^{\prime}$. We show our frequency perturbation module in Eq. \ref{eq: Frequency Perturbation Module} as follows: 
\begin{equation}
   \bm{\mathrm{x}} \mapsto \mathcal{D}(\bm{\mathrm{x}})  \mapsto \underbrace{\varphi(\mathcal{D}(\bm{\mathrm{x}}),\bm{\mathrm{q}})} _{\substack{\text{quantization,}\\ \text{rounding and}\\\text{de-quantization}}}\mapsto \mathcal{D}_{I}(\varphi(\cdot)) \mapsto \bm{\mathrm{x}}^{\prime}
   \label{eq: Frequency Perturbation Module}
\end{equation}
\noindent We repeat the above mentioned sequence of transformations for all patches and then merge  $\{\bm{\mathrm{x}}_i^{\prime}\}_{i=1}^{n}$ to form adversarial image $\mathrm{X}^{\prime} \in \mathbb{R}^{H\times W\times D}$. %

\begin{algorithm}[!t]
    \centering
    \caption{Volumetric Adversarial Frequency Attack (\textbf{VAFA}) }\label{alg:afa-algo}
    \footnotesize
    \begin{algorithmic}[1]
        \State Number of Steps: $T$,~~Quantization Threshold: $ q_{\text{max}}$
        \State \textbf{Input:} $\mathrm{X}\in\mathbb{R}^{H\times W\times D},\mathrm{Y}\in\{0,1\}^{\mathrm{NumClass}\times H\times W\times D}~~~~\textbf{Output:~} \mathrm{X}^{\prime}\in\mathbb{R}^{H\times W\times D}$ 
        \Function{{\textbf{VAFA}}}{X,Y}
             \State $\bm{\mathrm{q}}_i \gets \bm{1}~~~~~\forall~i\in\{1,2,\dots,n\}$ \Comment{Initialize all \textit{quantization tables} with ones.~~~~}
             \For{$t \gets 1$ to $T$}
                 \State $ \{\bm{\mathrm{x}}_i\}_{i=1}^{n} \gets \text{Split}(\mathrm{X})$~\Comment{Split $\mathrm{X}$ into 3D patches of size $(h\times w \times d)~~$}
                 \State $\bm{\mathrm{x}}_i^{\prime} \gets \mathcal{D}_{I}\big(~\varphi\bm{(}\mathcal{D}(\bm{\mathrm{x}}_i),\bm{\mathrm{q}}_{i}\bm{)}~\big)~~~\forall~i\in\{1,2,\dots,n\}$ \Comment{Frequency Perturbation~~~}
                 \State $ \mathrm{X}^{\prime} \gets \text{Merge}(\{\bm{\mathrm{x}}_i^{\prime}\}_{i=1}^{n})$~\Comment{Merge all adversarial patches to form $\mathrm{X}^{\prime}~~~~~$}
                 \State $\mathcal{L}( {\mathrm{X}}, {\mathrm{X}}^{\prime}, {\mathrm{Y}} ) = \mathcal{L}_{\mathrm{dice}} (\mathcal{M}_{\theta}({\mathrm{X}}^{\prime}), {\mathrm{Y}}) - \mathcal{L}_{\mathrm{ssim}}({\mathrm{X}},{\mathrm{X}}^{\prime})$
                 \State $\bm{\mathrm{q}}_{i} \gets \bm{\mathrm{q}}_{i} +~\mathrm{sign}(\nabla_{\bm{\mathrm{q}}_{i}} \mathcal{L}) ~~~~~\forall~i\in\{1,2,\dots,n\}$
                 \State $\bm{\mathrm{q}}_{i} \gets \mathrm{clip}(\bm{\mathrm{q}}_{i},~ \text{min=1},~\text{max=}q_{\text{max}})~~~~~\forall~i\in\{1,2,\dots,n\}$
             \EndFor
        \EndFunction
        \State \textbf{Return}  $\mathrm{X}^{\prime}$
    \end{algorithmic}
\end{algorithm}

\noindent \textbf{Quantization Constraint:} %
We learn quantization table $\bm{\mathrm{q}}$ by maximizing the $\mathcal{L}_{\mathrm{dice}}$ while ensuring that $\|\bm{\mathrm{q}}\|_{\infty} \le q_{\text{max}}$. Quantization threshold $q_{\text{max}}$ controls the extent to which DCT coefficients are perturbed. The higher the value of $q_{\text{max}}$, the more information is lost. The drop in perception quality of the adversarial sample and the accuracy of the model are directly proportional to the value of $q_{\text{max}}$. To increase the perceptual quality of adversarial samples, we also minimize the structural similarity loss \cite{wang2004image} between clean and adversarial samples, denoted by $\mathcal{L}_{\text{ssim}}({\mathrm{X}},{\mathrm{X}}^{\prime})$, in optimization objective. Our attack optimizes the following objective to fool a target model $\mathcal{M}_{\theta}$:
\begin{equation} \label{eq: afa objective}
\begin{gathered}
\underset{ \bm{\mathrm{q}} }{\mathrm{maximize}}~~ \mathcal{L}_{\mathrm{dice}} (\mathcal{M}_{\theta}({\mathrm{X}}^{\prime}), {\mathrm{Y}}) - \mathcal{L}_{\mathrm{ssim}}({\mathrm{X}},{\mathrm{X}}^{\prime}) \\
\mathrm{s.t.}~~ \|\bm{\mathrm{q}}\|_{\infty} \le q_{\mathrm{max}},
\end{gathered}
\end{equation}
where $\mathcal{L}_{\mathrm{ssim}}({\mathrm{X}},{\mathrm{X}}^{\prime}) = 1-\frac{1}{n} \sum_{i=1}^{n} \mathrm{SSIM}(\bm{\mathrm{x}}_i,\bm{\mathrm{x}}^{\prime}_i) $ is structural similarity loss \cite{wang2004image}. Algorithm \ref{alg:afa-algo} presents our volumetric adversarial frequency attack (VAFA). An overview of the attack can be found in \textit{maximization} step of Fig. \ref{fig: main-figure}.

\begin{algorithm}[!t]
    \centering
    \caption{Volumetric Adversarial Frequency Training (\textbf{VAFT})}\label{alg:afa-at}
    \footnotesize
    \begin{algorithmic}[1]
        \State Train Dataset: $\mathcal{X}=\{({\mathrm{X}}_i,{\mathrm{Y}}_i)\}_{i=1}^{N}$,~~${\mathrm{X}}_i \in \mathbb{R}^{H\times W\times D}$,~~${\mathrm{Y}}_i \in \{0,1\}^{\mathrm{NumClass}\times H\times W\times D}$
        \State NumSamples=$N$, BatchSize=$B$, Target Model: $\mathcal{M}_{\theta}$,~AT Robust Model: $\mathcal{M}_{_{\text{\faShield*}}}$
        \For{$i\gets 1~\text{to}~ \text{NumEpochs}$}
        \For{$j\gets 1~\text{to}~ \lfloor{N/B}\rfloor$}
        \State Sample a mini-batch $\mathcal{B} \subseteq \mathcal{X}$ of size $B$
        \State ${\mathrm{X}}^{\prime} \gets \textbf{VAFA}({\mathrm{X}},{\mathrm{Y}})~~\forall ({\mathrm{X}},{\mathrm{Y}})\in\mathcal{B}$ \Comment{Adv. Freq. Attack on clean images.}
        \State $\mathcal{L} = \mathcal{L}_{\text{dice}} (\mathcal{M}_{\theta}({\mathrm{X}}), {\mathrm{Y}})+ \mathcal{L}_{\mathrm{dice}}(\mathcal{M}_{\theta}({\mathrm{X}^{\prime}}),{\mathrm{Y}})+ \mathcal{L}_{\mathrm{fr}}(\mathcal{M}_{\theta}({\mathrm{X}}),\mathcal{M}_{\theta}({\mathrm{X}}^{\prime}))$
        \State Backward pass and update $\mathcal{M}_{\theta}$
        \EndFor
        \EndFor
        \State  $\mathcal{M}_{\Scale[0.4]{{{\text{\faShield*}}}}} \gets \mathcal{M}_{\theta}$ \Comment{AT robust model after training completion.}
        \State \textbf{Return} $\mathcal{M}_{_{{\text{\faShield*}}}}$
    \end{algorithmic}
\end{algorithm}

\subsection{Volumetric Adversarial Frequency Training (VAFT)}
\label{subsec: Adversarial Frequency Training}
The model parameters are then updated by minimizing the segmentation loss on both clean and adversarial samples (Eq. \ref{eq: adversarial frequency training}). Since our attack disrupts the frequency domain to find adversaries, we develop a novel frequency consistency loss (Eq. \ref{eq: fr loss}) to encourage frequency domain representation of the model's output (segmentation logits) for the clean sample close to the adversarial sample. Our frequency consistency loss not only boosts the robustness of the model against adversarial attacks but also improves/retains the performance of the robust model on clean images (Sec. \ref{sec: Experiments and Results}). We present our volumetric adversarial frequency training (VAFT) in Algo. \ref{alg:afa-at}. %
\begin{equation}\label{eq: adversarial frequency training}
 \underset{ \theta }{\mathrm{minimize}}~ \mathcal{L}_{\text{dice}} (\mathcal{M}_{\theta}({\mathrm{X}}), {\mathrm{Y}})+  \mathcal{L}_{\text{dice}} (\mathcal{M}_{\theta}({\mathrm{X}}^{\prime}), {\mathrm{Y}}) + \mathcal{L}_{_{\mathrm{fr}}}(\mathcal{M}_{\theta}({\mathrm{X}}),\mathcal{M}_{\theta}({\mathrm{X}}^{\prime})),   
\end{equation}
\begin{equation} 
\label{eq: fr loss}
\mathcal{L}_{_{\mathrm{fr}}}(\mathcal{M}_{\theta}({\mathrm{X}}),\mathcal{M}_{\theta}({\mathrm{X}}^{\prime})) = \|\mathcal{D}(\mathcal{M}_{\theta}({\mathrm{X}}))-\mathcal{D}(\mathcal{M}_{\theta}({\mathrm{X}}^{\prime}))\|_{_1},
\end{equation}
where ${\mathrm{X}}^{\prime} = \textbf{VAFA}(\mathrm{X},\mathrm{Y})$ and $\mathcal{D}(\cdot)$ is 3D DCT function. An overview of the adversarial training can be found in \textit{minimization} step of Fig. \ref{fig: main-figure}.\\

\noindent Fig. \ref{fig: visualization} presents a qualitative results of adversarial examples under different attacks on the standard UNETR model. We highlight areas by red bounding box in Fig. \ref{fig: visualization} to show the impact of each attack on the model performance, when compared with prediction on clean sample. Our attack (VAFA) achieves higher fooling rate as compared to other voxel-domain attacks, while maintaining comparable perceptual similarity.

\section{Experiments and Results}
\label{sec: Experiments and Results}

\textbf{Implementation Details:} We demonstrate the effectiveness of our approach using two medical segmentation models: UNETR\cite{hatamizadeh2022unetr}, UNETR++\cite{shaker2022unetr++} and two datasets: Synapse (18-12 split) \cite{landman2015miccai}, and ACDC \cite{bernard2018deep}. Using pre-trained models from open-source Github repositories by the corresponding authors, we launch different adversarial attacks and conduct adversarial training with default parameters. We use the Pytorch framework and single NVIDIA A100-SXM4-40GB GPU for our experiments. For a pixel/voxel range $[0,255]$, we create $l_{\infty}$ adversarial examples under perturbation budgets of $\epsilon \in \{4,8\}$  for voxel-domain attacks following \cite{duan2021advdrop} and compare it with our attack \textbf{VAFA}. Unless otherwise specified, all attacks are run for a total of $20$ optimization steps. More details about the parameters of the attacks used in different experiments can be found in Appendix. We use mean Dice Similarity Score $(\mathrm{DSC})$, mean 95\% Hausdorff Distance $(\mathrm{HD95})$. We also report perceptual similarity between clean and adversarial sample $(\mathrm{LPIPS})$ \cite{zhang2018perceptual}.%

\begin{table}
\centering
\begin{minipage}{0.22\textwidth}
\caption{\small{Voxel vs. Freq. Attacks}}
\label{tbl:afa3d-afa2d-voxel-attacks}
\scalebox{0.8}{
\begin{tabular}{ l | c  c } 
\toprule
\rowcolor{Gray}
 Attack & $\mathrm{DSC}\hspace{-0.3em}\downarrow$ & $\mathrm{LPIPS}\hspace{-0.3em}\uparrow$ \\
\midrule
~~~-  & \textcolor{OliveGreen}{74.31} & - \\
 PGD    & 62.67 &  \textbf{98.94} \\ 
 FGSM   & 62.77 &  98.82 \\ 
 BIM    & 62.76 &  98.93 \\ 
 GN     & 74.19 &  97.71 \\
 \hline
VAFA-2D & 61.66 &  98.43 \\ 
 \rowcolor{lightskyblue}
 VAFA & \textbf{52.54} & 97.84 \\
 \bottomrule
\end{tabular}
}
\end{minipage}
\hspace{1.1em}
\begin{minipage}{0.21\textwidth}
\caption{Impact of $q_{\text{max}}$ on VAFA}
\label{tbl:afa-qmax-nossim-ssim}
\scalebox{0.8}{
\begin{tabular}{ l | c  c } 
\toprule
  \rowcolor{Gray}
  $q_{\text{max}}$ & $\mathrm{DSC}\hspace{-0.3em}\downarrow$ & $\mathrm{LPIPS}\hspace{-0.3em}\uparrow$ \\
  \midrule
   ~-   & \textcolor{OliveGreen}{74.31} & -  \\
   10   & 65.95 & \textbf{99.10} \\ 
   20   & 56.24 & 98.70 \\ 
   30   & 50.96 & 98.33 \\ 
   40   & 49.58 & 97.90\\
   60   & 48.83 & 96.60 \\
   80   & \textbf{48.76} & 94.50 \\
   \bottomrule
\end{tabular}
}
\end{minipage}
\hspace{0.1em}
\begin{minipage}{0.21\textwidth}
\caption{Impact of steps on VAFA}
\label{tbl:afa-steps}
\scalebox{0.8}{
\begin{tabular}{ c | c c} 
\toprule
\rowcolor{Gray}
Steps & $\mathrm{DSC}\hspace{-0.3em}\downarrow$ & $\mathrm{LPIPS}\hspace{-0.3em}\uparrow$ \\
  \midrule
   -  & \textcolor{OliveGreen}{74.31} & - \\
   10   & 61.33 & \textbf{98.85} \\ 
   20   & 56.24 &  98.70 \\
   30   & 54.37 &  98.64 \\ 
   40   & 53.31 &  98.59 \\
   50   & 52.97 & 98.54 \\
   60   & \textbf{52.25} & 98.52 \\
   \bottomrule
\end{tabular}
}
\end{minipage}
\hspace{0.1em}
\begin{minipage}{0.23\textwidth}
\caption{Impact of patch size on VAFA}
\label{tbl:afa-patchsize}
\scalebox{0.8}{
\begin{tabular}{ l | c  c }
\toprule
\rowcolor{Gray}
 {Size} & ~$\mathrm{DSC}\hspace{-0.3em}\downarrow$~ & $\mathrm{LPIPS}\hspace{-0.3em}\uparrow$\\
\midrule
-  & \textcolor{OliveGreen}{74.31} & -  \\
 $4$     & 63.48  & \textbf{98.90} \\ 
 $8$     & 56.24  & 98.70 \\
 $16$    & 41.30  & 98.14 \\
 $32$    & 32.40  & 97.49 \\
 $48$    & 28.19  & 97.16 \\
 $96$    & \textbf{28.08} & 96.47 \\
\bottomrule
\end{tabular}
}
\end{minipage}
\end{table}

\noindent\textbf{Results:} For each evaluation metric, we take mean across all classes (including background) and test images. In each table (where applicable), \textcolor{OliveGreen}{green values} show $\mathrm{DSC}$ and $\mathrm{HD95}$ on clean images.
Table \ref{tbl:afa3d-afa2d-voxel-attacks} shows comparison of voxel-domain attacks (e.g. PGD \cite{madry2017towards}, FGSM \cite{goodfellow2014explaining}, BIM \cite{kurakin2018adversarial}, GaussianNoise(GN) \cite{kim2020torchattacks}) with VAFA-2D (2D DCT in FPM applied on each scan independently) and VAFA on UNETR model (Synapse). VAFA achieves a higher fooling rate as compared to other attacks with comparable LPIPS. We posit that VAFA-2D on volumetric MIS data is sub-optimal and it does not take into account the 3D nature of the data and model\textquotesingle s reliance on the 3D neighborhood of a voxel to predict its class.  Further details are provided in the supplementary material. We show impacts of different parameters of VAFA e.g. quantization threshold $(q_{\text{max}})$, steps, and patch size $(h\times w \times d)$ on $\mathrm{DSC}$ and $\mathrm{LPIPS}$ in Table. \ref{tbl:afa-qmax-nossim-ssim},\ref{tbl:afa-steps} and \ref{tbl:afa-patchsize} respectively. $\mathrm{DSC}$ and $\mathrm{LPIPS}$ decrease when these parameters values are increased. Table \ref{tbl:unetr-unetrpp-pix-vs-freq} shows a comparison of VAFA (patch size = $32\times 32 \times 32$) with other voxel-domain attacks on UNETR and UNETR++ models. For adversarial training experiments, we use $q_{\text{max}}=20$ (for Synapse), $q_{\text{max}}=10$ (for ACDC) and patch-size of $32\times32\times32$ (chosen after considering the trade-off between DSC and LPIPS from Table \ref{tbl:afa-patchsize}) for VAFA.  For voxel-domain attacks, we use $\epsilon=4$ (for Synapse) and $\epsilon=2$ (for ACDC)  by following the work of \cite{guo2017countering,prakash2018deflecting}. Table \ref{tbl:unetr-unetrpp-at} presents a comparison of the performance ($\mathrm{DSC}$) of various adversarially trained models against different attacks. %
$\mathcal{M}^{\Scale[0.5]{\text{VAFA-FR}}}_{_{\text{\faShield*}}}$, $\mathcal{M}^{\Scale[0.5]{\text{VAFA}}}_{_{\text{\faShield*}}}$ denote our robust models which were adversarially trained with and without frequency consistency loss $(\mathcal{L}_{\text{fr}},~\text{Eq.}~\ref{eq: fr loss})$ respectively. In contrast to other voxel-domain robust models, our approach demonstrated robustness against both voxel and frequency-based attacks.

\begin{table}[!t]
\caption{Comparison of VAFA with other voxel-domain attacks (Synapse dataset).}
\centering \small
\setlength{\tabcolsep}{8pt}
\scalebox{0.75}[0.75]{
\begin{tabular}{ l | c  c  c | c  c  c} 
\toprule
\rowcolor{Gray}
 {Models $\rightarrow$} & \multicolumn{3}{c|}{\textbf{UNETR}} & \multicolumn{3}{c}{\textbf{UNETR++}}\\
 \rowcolor{Gray}
 Attacks $\downarrow$ & ~$\mathrm{DSC}\hspace{-0.3em}\downarrow$~ & $\mathrm{HD95}\hspace{-0.3em}\uparrow$~  &  $\mathrm{LPIPS}\hspace{-0.3em}\uparrow$~ & ~$\mathrm{DSC}\hspace{-0.3em}\downarrow$~ & $\mathrm{HD95}\hspace{-0.3em}\uparrow$~  &  $\mathrm{LPIPS}\hspace{-0.3em}\uparrow$ \\
 \midrule
 \textcolor{OliveGreen}{Clean Images}  & \textcolor{OliveGreen}{74.3} & \textcolor{OliveGreen}{14.0} & - &  \textcolor{OliveGreen}{84.7} & \textcolor{OliveGreen}{12.7} & - \\
 PGD\phantom{ii} {$(\epsilon=4/8)$}   & {62.7/50.8} & {40.4/64.5} & {\textbf{98.9}/95.3} & {77.5/67.1} & {48.1/78.3} & {95.7/85.1} \\
 FGSM {$(\epsilon=4/8)$}              & {62.8/53.9} & {34.8/48.7} & {98.8/94.7} & {73.1/67.1} & {37.3/43.2} & {94.7/82.2} \\
 BIM\phantom{M}  {$(\epsilon=4/8)$}   & {62.8/50.7} & {39.9/\textbf{65.8}} & {98.8/95.3} & {77.3/66.8} & {46.6/78.1} & {\textbf{95.8}/85.3} \\
 GN\phantom{SM} {$(\sigma=4/8)$}      & {74.2/73.9} & {17.0/15.4} & {97.7/91.1} & {84.7/84.3} & {12.3/13.4} & {93.3/78.2} \\
 \rowcolor{lightskyblue}
 VAFA {$(q_{\text{max}}=20/30)$}       & {\textbf{32.2}/\textbf{29.8}} & {\textbf{57.6}/59.9} & {97.5/\textbf{96.9}} & {\textbf{45.3}/\textbf{39.3}} & {\textbf{73.9}/\textbf{85.2}} & {94.2/\textbf{94.7}} \\
\bottomrule
\end{tabular}
}
\label{tbl:unetr-unetrpp-pix-vs-freq}
\end{table}

\begin{table}[!t]

\caption{Performance of different attacks on adversarially trained (robust) models. }
\label{tbl:unetr-unetrpp-at}
\centering
\small
\setlength{\tabcolsep}{8pt}
\scalebox{0.7}[0.7]{
\begin{tabular}{l | l | c  c  c  c  c | c  c  c  c  c } 
\toprule
 \rowcolor{Gray}
 {} & {} & \multicolumn{5}{c|}{\textbf{UNETR}} & \multicolumn{5}{c}{\textbf{UNETR++}} \\
 \rowcolor{Gray}
 {} & \multirow{-2}{*}{\makecell[l]{ \text{Attacks $\rightarrow$} \\ \text{Models $\downarrow$} }} & Clean & PGD & FGSM & BIM & VAFA & Clean & PGD & FGSM & BIM & VAFA \\
 \midrule
 \multirow{6}{*}{\rotatebox[origin=c]{90}{\parbox[c]{1.5cm}{\centering\texttt{Synapse}}}} & $\mathcal{M}^{\Scale[0.5]{\text{PGD}}}_{_{\text{\faShield*}}}$     & 73.47 & 65.53 &  65.68  &  65.51 &  42.47 & 75.43 & 67.81 & 67.82 & 67.80 & 38.22  \\
 &$\mathcal{M}^{\Scale[0.5]{\text{FGSM}}}_{_{\text{\faShield*}}}$        & 72.44 & 64.80 &  66.31  &  64.76 &  39.02  & 81.06 & 73.84 & 74.76 & 73.77 & 37.48  \\
 &$\mathcal{M}^{\Scale[0.5]{\text{BIM}}}_{_{\text{\faShield*}}}$         & 75.12 & 67.78 &  68.32  &  67.78 &  45.97 & 74.80 & 67.58 & 67.46 & 67.57 & 35.72   \\
 &$\mathcal{M}^{\Scale[0.5]{\text{GN}}}_{_{\text{\faShield*}}}$          & 73.17 & 61.40 &  61.77  &  61.29 &  30.00 & 80.05 & 76.23 & 70.96 & 74.51 & 41.44   \\
 &$\mathcal{M}^{\Scale[0.5]{\text{VAFA}}}_{_{\text{\faShield*}}}$        & 74.67 & 64.83 &  65.49  &  64.73 &  66.31 & 81.88 & 69.09 & 65.40 & 68.90 & 76.47   \\
 \rowcolor{lightskyblue}
 \cellcolor{white} & $\mathcal{M}^{\Scale[0.5]{\text{VAFA-FR}}}_{_{\text{\faShield*}}}$ & \textbf{75.66} & 65.90  &  66.79 &  65.83 & 66.33 & \textbf{82.65} & 70.61 & 67.00 & 70.41 & 78.19  \\
 \midrule
   {} & $\mathcal{M}^{\Scale[0.5]{\text{VAFA}}}_{_{\text{\faShield*}}}$ & 81.95 & 60.77 & 68.16 & 60.75 & 69.76 & 89.00 & 76.28 & 80.41 & 76.56 & 88.45 \\
  \rowcolor{lightskyblue}
  \cellcolor{white}  \multirow{-2}{*}{\rotatebox[origin=c]{90}{\parbox[c]{0.8cm}{\centering\texttt{ACDC}}}} & $\mathcal{M}^{\Scale[0.5]{\text{VAFA-FR}}}_{_{\text{\faShield*}}}$ & \textbf{83.44} & 60.63 & 69.33 & 60.61 & 73.05 & \textbf{91.36} & 85.42 & 87.42 & 83.90 & 91.23 \\
 \bottomrule
\end{tabular}
}
\end{table}

\section{Conclusion}
We present a frequency-domain based adversarial attack and training for volumetric medical image segmentation. Our attack strategy is tailored to the 3D nature of medical imaging data, allowing for a higher fooling rate than voxel-based attacks while preserving comparable perceptual similarity of adversarial samples. Based upon our proposed attack, we introduce a frequency-domain adversarial training method that enhances the robustness of the volumetric  segmentation model against both voxel and frequency-domain based attacks. Our training strategy is particularly important in medical image segmentation, where the accuracy and reliability of the model are crucial for clinical decision making.

\bibliographystyle{splncs04}
\bibliography{main}

\end{document}


\title{Frequency Domain Adversarial Training for Robust Volumetric Medical Segmentation}

\author{Supplementary Material}
\institute{}
\maketitle %

\begin{table}
\centering
\begin{minipage}{1\textwidth}
\centering
\caption{\small{Settings of different adversarial attacks used in Table [1-6] of \textbf{main manuscript}. First column shows the Table \# (used in the \textbf{main manuscript}) and corresponding row shows the settings of the adversarial attacks used in that table.}}
\label{tbl:afa3d-afa2d-voxel-attacks}
\scalebox{0.75}{
\begin{tabular}{ l | l  c} 
\toprule
\rowcolor{Gray}
 \textbf{Table} \#~~ & \textbf{Attack Settings} &  \\
\midrule
 Table-1 & \makecell[l]{\texttt{PGD, FGSM, BIM, GN}: \big(Steps=20, StepSize=$3$, $\epsilon=4$\big)\\
 \texttt{VAFA}: \big(Steps=20, $q_{\text{max}}=20$,  PatchSize(for 2D case)=$(8\times 8)$, PatchSize(for 3D case)=$(8\times 8\times 8)$\big)}  \\ \hline
 Table-2 &  \texttt{VAFA}: \big(Steps=20, $q_{\text{max}}\in\{10,20,30,40,60,80\}$, PatchSize=$(8\times 8\times 8)$, $\mathcal{L}=\mathcal{L}_{\text{dice}}-\mathcal{L}_{\text{ssim}}$\big)\\ \hline 
 Table-3 &  \texttt{VAFA}: \big($\text{Steps}\in\{10,20,30,40,50,60\}$, $q_{\text{max}}=20$, PatchSize=$(8\times 8\times 8)$, $\mathcal{L}=\mathcal{L}_{\text{dice}}-\mathcal{L}_{\text{ssim}}$\big)\\ \hline 
 Table-4 &  \texttt{VAFA}: \big(Steps=20, $q_{\text{max}}=20$, $\text{PatchSize}=(i\times i\times i)~\forall i \in \{4,8,16,32,48,96\}$, $\mathcal{L}=\mathcal{L}_{\text{dice}}-\mathcal{L}_{\text{ssim}}$\big)\\ \hline 
 Table-5 & \makecell[l]{\texttt{PGD, FGSM, BIM, GN}: \big(Steps=20, StepSize=$3$, $\epsilon\in\{4,8\}$\big)\\
 \texttt{VAFA}: \big(Steps=20, $q_{\text{max}}=\{20,30\}$, PatchSize=$(32\times 32\times 32)$, $\mathcal{L}=\mathcal{L}_{\text{dice}}-\mathcal{L}_{\text{ssim}}$\big)} \\ \hline
 Table-6 &  \makecell[l]{Synapse Dataset: \texttt{PGD, FGSM, BIM}: \big(Steps=20, StepSize=$3$, $\epsilon=4$\big)\\
 Synapse Dataset: \texttt{VAFA}: \big(Steps=20, $q_{\text{max}}=20$, PatchSize=$(32\times 32\times 32)$, $\mathcal{L}=\mathcal{L}_{\text{dice}}-\mathcal{L}_{\text{ssim}}$\big)\\
 ACDC Dataset~~: \texttt{PGD, FGSM, BIM}: \big(Steps=20, StepSize=$3$, $\epsilon=2$\big)\\
 ACDC Dataset~~: \texttt{VAFA}: \big(Steps=20, $q_{\text{max}}=10$, PatchSize=$(32\times 32\times 32)$, $\mathcal{L}=\mathcal{L}_{\text{dice}}-\mathcal{L}_{\text{ssim}}$\big)}\\
 \bottomrule
\end{tabular}
}
\end{minipage}
\vspace{-3em}
\end{table}

\begin{table}
\begin{minipage}{0.30\textwidth}
\caption{Voxel vs. Freq. Domain Attacks}
\label{tbl:afa3d-afa2d-voxel-attacks-supp}
\scalebox{0.8}{
\begin{tabular}{ l | c c c } 
\toprule
\rowcolor{Gray}
 Attack & $\mathrm{DSC}\hspace{-0.3em}\downarrow$ & $\mathrm{HD95}\hspace{-0.3em}\uparrow$ & $\mathrm{LPIPS}\hspace{-0.3em}\uparrow$ \\
\midrule
~~~-  & \textcolor{OliveGreen}{74.31} & \textcolor{OliveGreen}{14.03} & - \\
 PGD    & 62.67 & 40.36 &  \textbf{98.94} \\ 
 FGSM   & 62.77 & 34.77 &  98.82 \\ 
 BIM    & 62.76 & 39.94 &  98.93 \\ 
 GN     & 74.19 & 16.99 &  97.71 \\
 \hline
VAFA-2D & 61.66 & 37.22 &  98.43 \\ 
 \rowcolor{lightskyblue}
 VAFA & \textbf{52.54} & \textbf{47.55} & 97.84 \\
 \bottomrule
\end{tabular}
}
\end{minipage}
\hspace{2.3em}
\begin{minipage}{0.27\textwidth}
\caption{Impact of $q_{\text{max}}$ on VAFA}
\label{tbl:afa-qmax-nossim-ssim}
\scalebox{0.8}{
\begin{tabular}{ l | c c c } 
\toprule
  \rowcolor{Gray}
  $q_{\text{max}}$ & $\mathrm{DSC}\hspace{-0.3em}\downarrow$ & $\mathrm{HD95}\hspace{-0.3em}\uparrow$ & $\mathrm{LPIPS}\hspace{-0.3em}\uparrow$ \\
  \midrule
   ~-   & \textcolor{OliveGreen}{74.31} & \textcolor{OliveGreen}{14.03} & -  \\
   10   & 65.95 & 26.25 & \textbf{99.10} \\ 
   20   & 56.24 & 35.92 & 98.70 \\ 
   30   & 50.96 & 44.09 & 98.33 \\ 
   40   & 49.58 & 43.66 & 97.90\\
   60   & 48.83 & 44.55 & 96.60 \\
   80   & \textbf{48.76} & \textbf{45.30} & 94.50 \\
   \bottomrule
\end{tabular}
}
\end{minipage}
\hspace{2.3em}
\begin{minipage}{0.27\textwidth}
\caption{Impact of steps on VAFA}
\label{tbl:afa-steps}
\scalebox{0.8}{
\begin{tabular}{ c | c c c} 
\toprule
\rowcolor{Gray}
Steps & $\mathrm{DSC}\hspace{-0.3em}\downarrow$ & $\mathrm{HD95}\hspace{-0.3em}\uparrow$ & $\mathrm{LPIPS}\hspace{-0.3em}\uparrow$ \\
  \midrule
   -  & \textcolor{OliveGreen}{74.31} & \textcolor{OliveGreen}{14.03} & - \\
   10   & 61.33 & 33.20 & \textbf{98.85} \\ 
   20   & 56.24 & 35.92 &  98.70 \\
   30   & 54.37 & 38.00 &  98.64 \\ 
   40   & 53.31 & 37.76 &  98.59 \\
   50   & 52.97 & \textbf{39.23} &  98.54 \\
   60   & \textbf{52.25} & 39.19 & 98.52 \\
   \bottomrule
\end{tabular}
}
\end{minipage}
\vspace{1.5em}
\\
\begin{minipage}{0.35\textwidth}
\caption{Impact of patch size on VAFA}
\label{tbl:afa-patchsize}
\scalebox{0.75}{
\begin{tabular}{ l | c c c }
\toprule
\rowcolor{Gray}
 {Patch Size} & ~$\mathrm{DSC}\hspace{-0.3em}\downarrow$~ & ~$\mathrm{HD95}\hspace{-0.3em}\uparrow$~ & $\mathrm{LPIPS}\hspace{-0.3em}\uparrow$\\
\midrule
~~~~~~~~~~-  & \textcolor{OliveGreen}{74.31} & \textcolor{OliveGreen}{14.03} & -  \\
 $(\phantom{0}4\times \phantom{0}4\times\phantom{0}4)$     & 63.48  & 32.63 & \textbf{98.90} \\ 
 $(\phantom{0}8\times \phantom{0}8\times\phantom{0}8)$     & 56.24  & 35.92 & 98.70 \\
 $(16 \times 16 \times 16)$    & 41.30  & 45.98 & 98.14 \\
 $(32 \times 32 \times 32)$    & 32.40  & 56.64 & 97.49 \\
 $(48 \times 48 \times 48)$    & 28.19  & \textbf{66.08} & 97.16 \\
 $(96 \times 96 \times 96)$    & \textbf{28.08} & 59.09 & 96.47 \\
\bottomrule
\end{tabular}
}
\end{minipage}
\hspace{1em}
\begin{minipage}{0.6\textwidth}
\centering
\includegraphics[height=2.7cm, width=1\textwidth, trim={0cm 0.1cm 0cm 0cm},clip ]{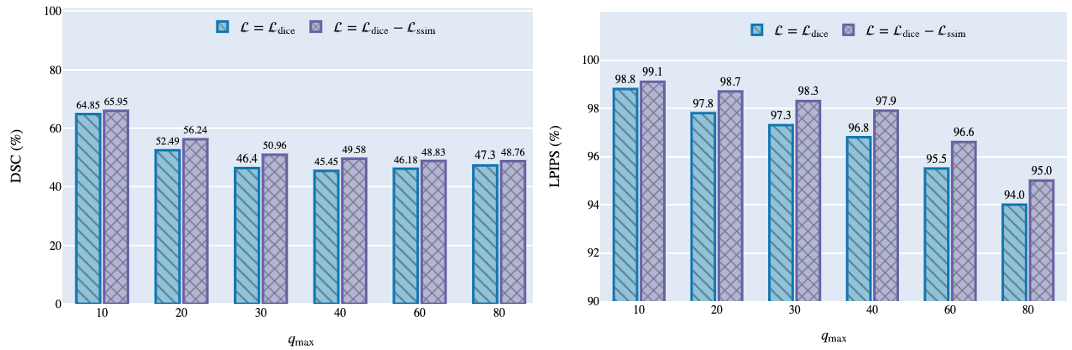}
\vspace{-1em}
\caption{Using SSIM loss $(\mathcal{L}_{\text{ssim}})$ in VAFA objective improves $\mathrm{LPIPS}$ and slightly increases $\mathrm{DSC}$.}
\end{minipage}
\end{table}

\begin{figure*}[!t]
\begin{minipage}{\textwidth}
\centering

\begin{minipage}{0.19\textwidth}
\centering
\tiny{\texttt{DSC=62.23~~~~\phantom{LPIPS=99.84}}}
\includegraphics[height=2.2cm, width=\linewidth,,  trim={1cm 6.5cm 1.5cm 1cm},clip ]{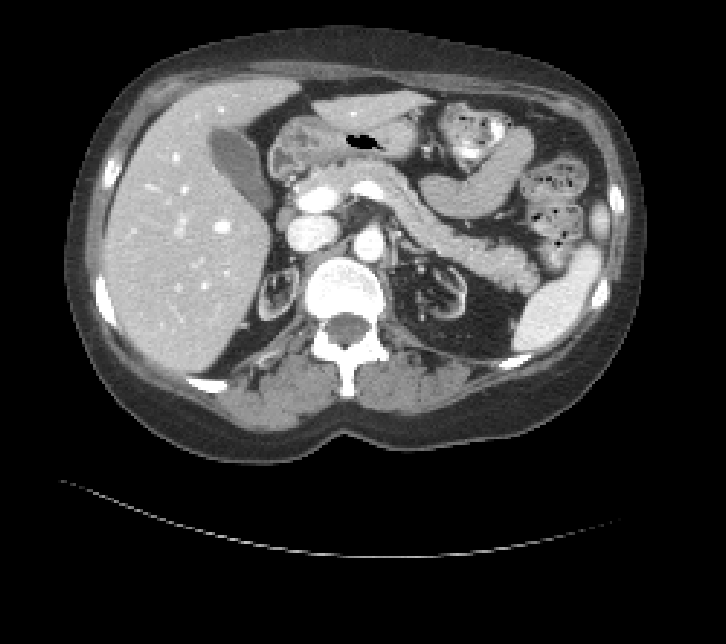}
\end{minipage}
\begin{minipage}{0.19\textwidth}
\centering
\tiny{\texttt{DSC=51.94~~~~LPIPS=99.16}}
\includegraphics[height=2.2cm, width=\linewidth,,  trim={1cm 6.5cm 1.5cm 1cm},clip ]{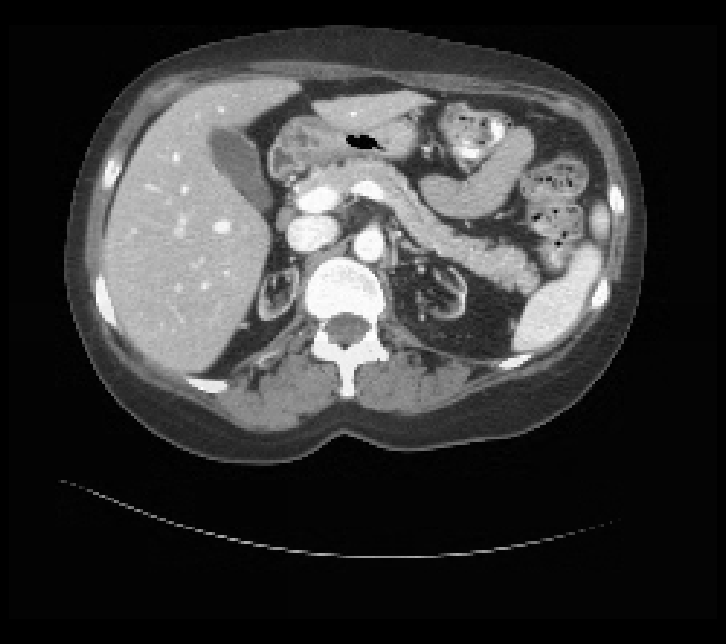}
\end{minipage}
\begin{minipage}{0.19\textwidth}
\centering
\tiny{\texttt{DSC=46.92~~~~LPIPS=98.25}}
\includegraphics[height=2.2cm, width=\linewidth,,  trim={1cm 6.5cm 1.5cm 1cm},clip ]{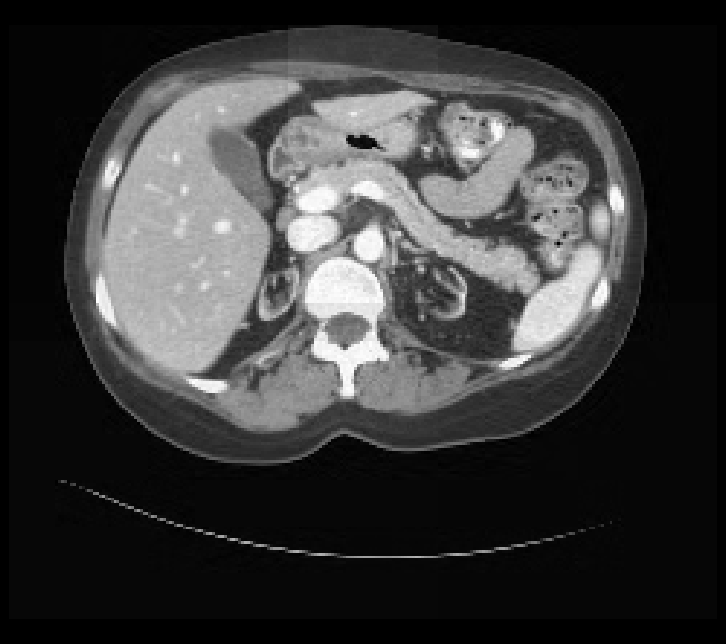}
\end{minipage}
\begin{minipage}{0.19\textwidth}
\centering
\tiny{\texttt{DSC=42.08~~~~LPIPS=97.02}}
\includegraphics[height=2.2cm, width=\linewidth,,  trim={1cm 6.5cm 1.5cm 1cm},clip ]{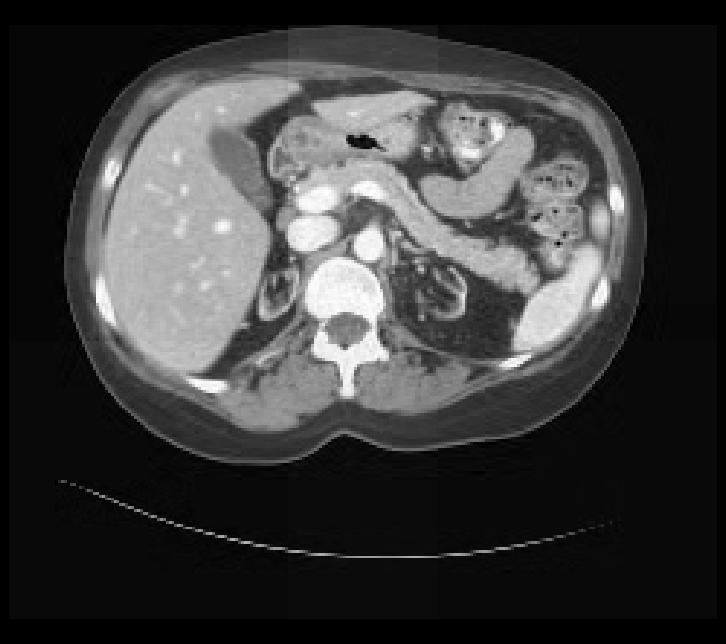}
\end{minipage}
\begin{minipage}{0.19\textwidth}
\centering
\tiny{\texttt{DSC=41.68~~~~LPIPS=95.59}}
\includegraphics[height=2.2cm, width=\linewidth,,  trim={1cm 6.5cm 1.5cm 1cm},clip ]{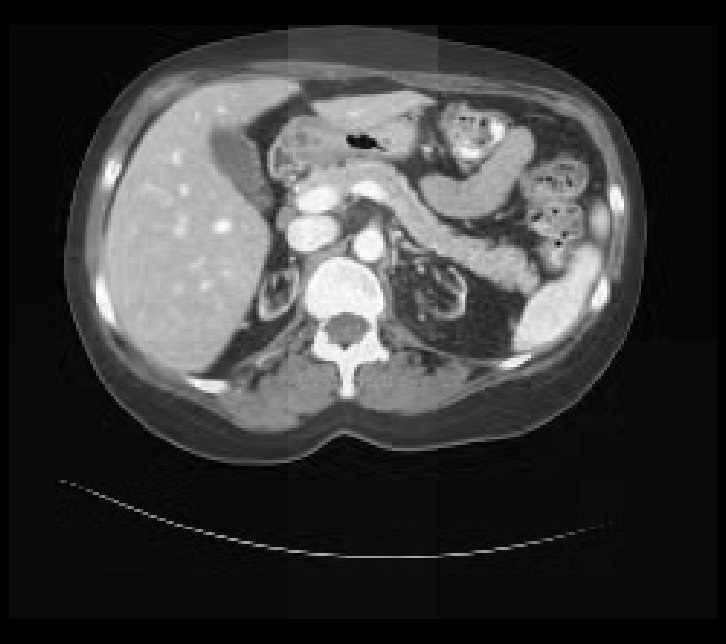}
\end{minipage}
\\
\begin{minipage}{0.19\textwidth}
\centering
\includegraphics[height=2.2cm, width=\linewidth,,  trim={1cm 6.5cm 1.5cm 1cm},clip ]{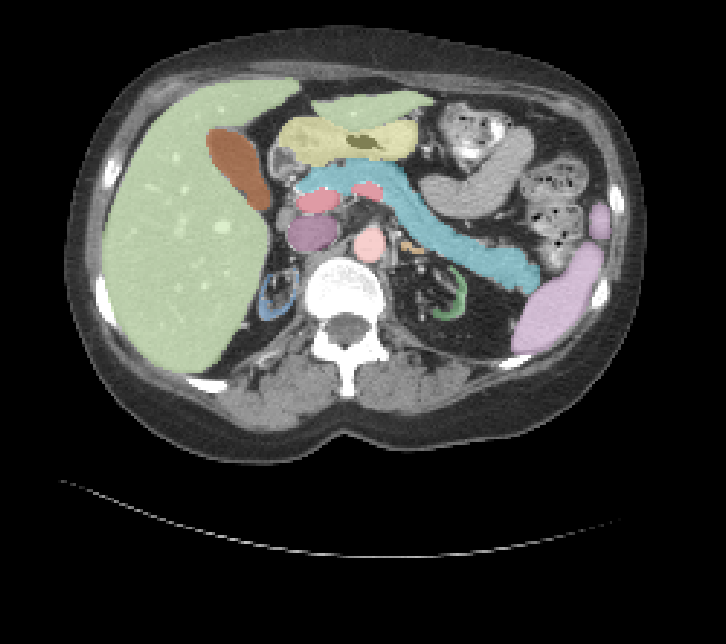}
\footnotesize $\mathrm{Clean}$
\end{minipage}
\begin{minipage}{0.19\textwidth}
\centering
\includegraphics[height=2.2cm, width=\linewidth,,  trim={1cm 6.5cm 1.5cm 1cm},clip ]{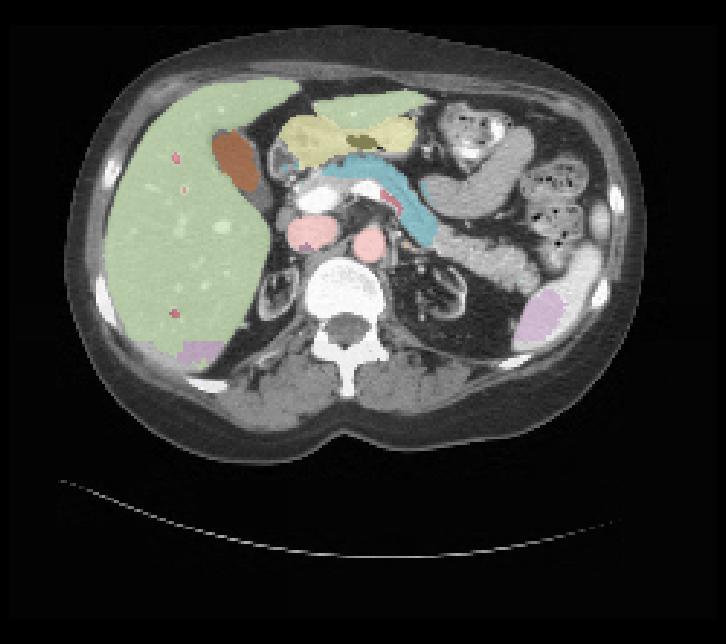}
\footnotesize $q_{\text{max}}=20$
\end{minipage}
\begin{minipage}{0.19\textwidth}
\centering
\includegraphics[height=2.2cm, width=\linewidth,,  trim={1cm 6.5cm 1.5cm 1cm},clip ]{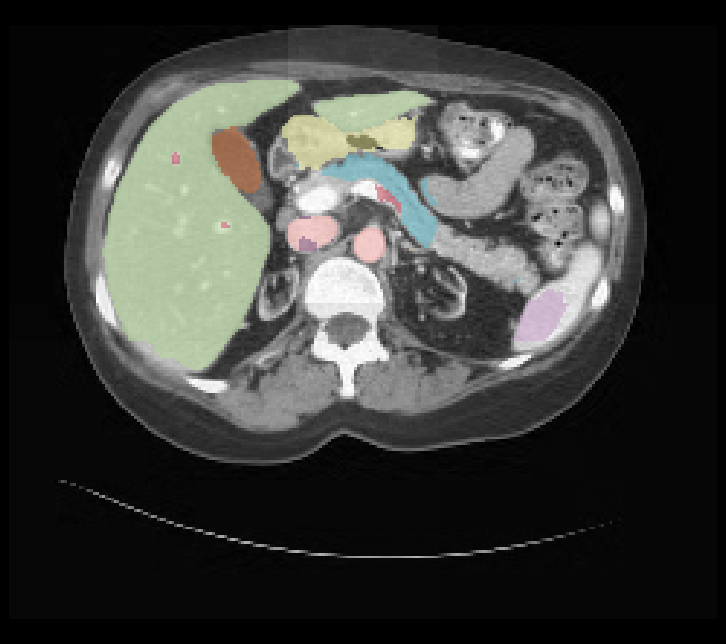}
\footnotesize $q_{\text{max}}=40$
\end{minipage}
\begin{minipage}{0.19\textwidth}
\centering
\includegraphics[height=2.2cm, width=\linewidth,,  trim={1cm 6.5cm 1.5cm 1cm},clip ]{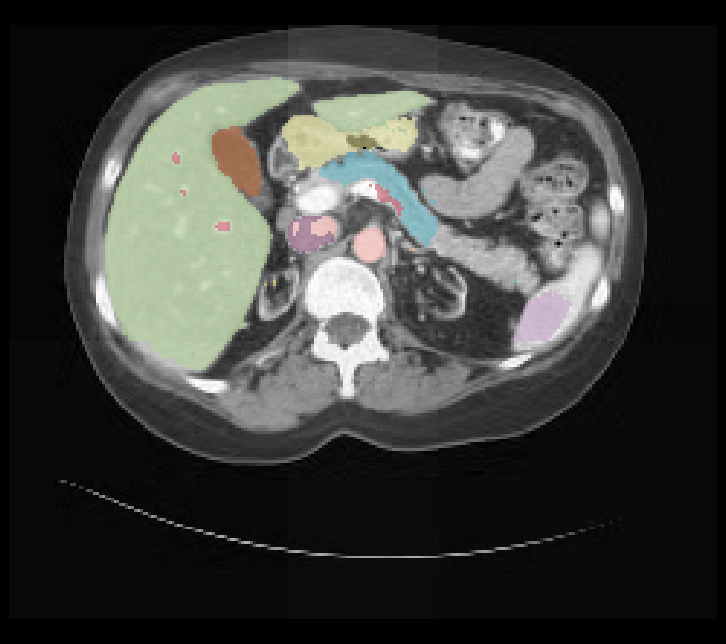}
\footnotesize $q_{\text{max}}=60$
\end{minipage}
\begin{minipage}{0.19\textwidth}
\centering
\includegraphics[height=2.2cm, width=\linewidth,,  trim={1cm 6.5cm 1.5cm 1cm},clip ]{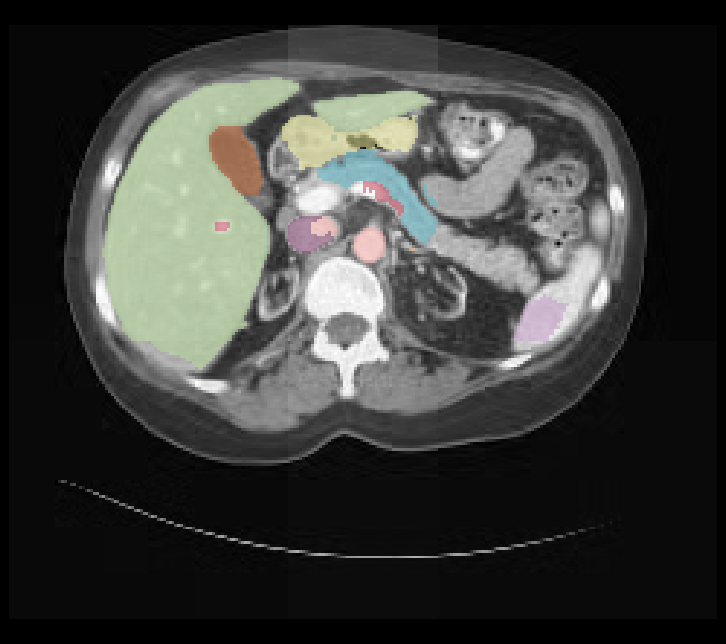}
\footnotesize $q_{\text{max}}=80$
\end{minipage}
\caption{Qualitative multi-organ segmentation results showing impact of quantization threshold $(q_{\text{max}})$ on model prediction in VAFA. Higher value of $q_{\text{max}}$ results in higher fooling rate (i.e. more reduction in Dice score). }
\label{fig:qmax-impact-qual}
\end{minipage}
\end{figure*}

\begin{figure*}[!t]

\begin{minipage}{\textwidth}
\centering
\rotatebox{90}{Synapse}
\begin{minipage}{0.25\textwidth}
\centering
\includegraphics[height=2.2cm, width=\linewidth, trim={1cm 4cm 0.5cm 8cm},clip ]{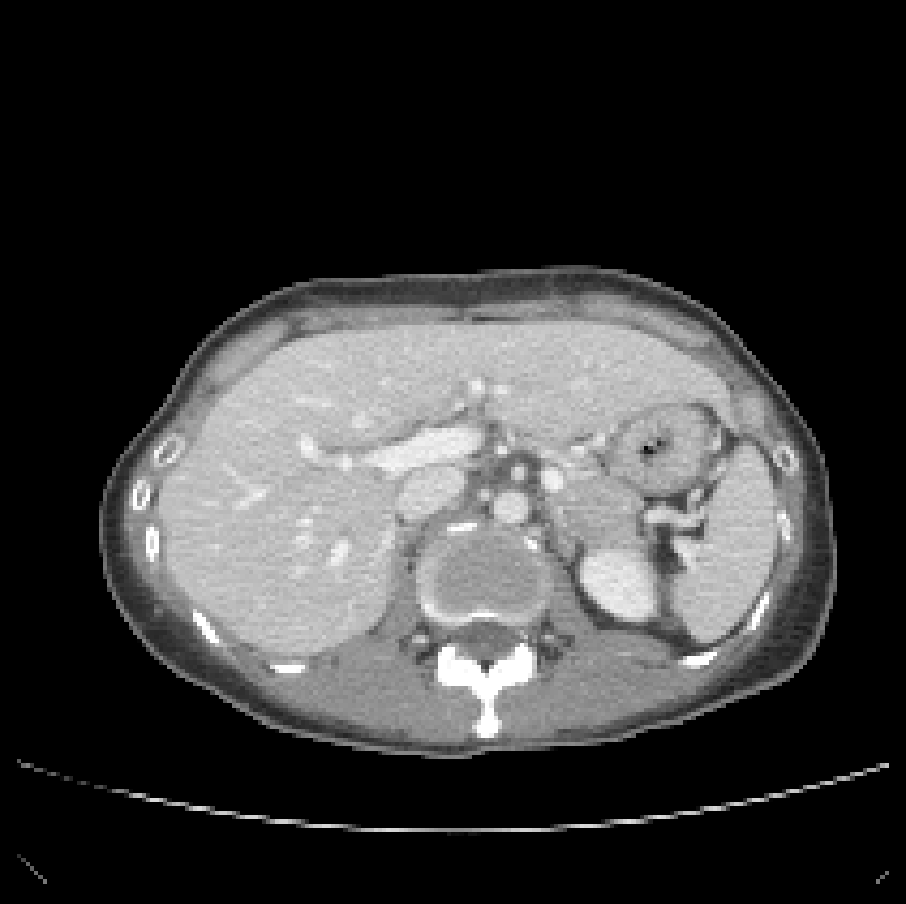}
\end{minipage}
\begin{minipage}{0.25\textwidth}
\centering
\includegraphics[height=2.2cm, width=\linewidth, trim={1cm 4cm 0.5cm 8cm},clip ]{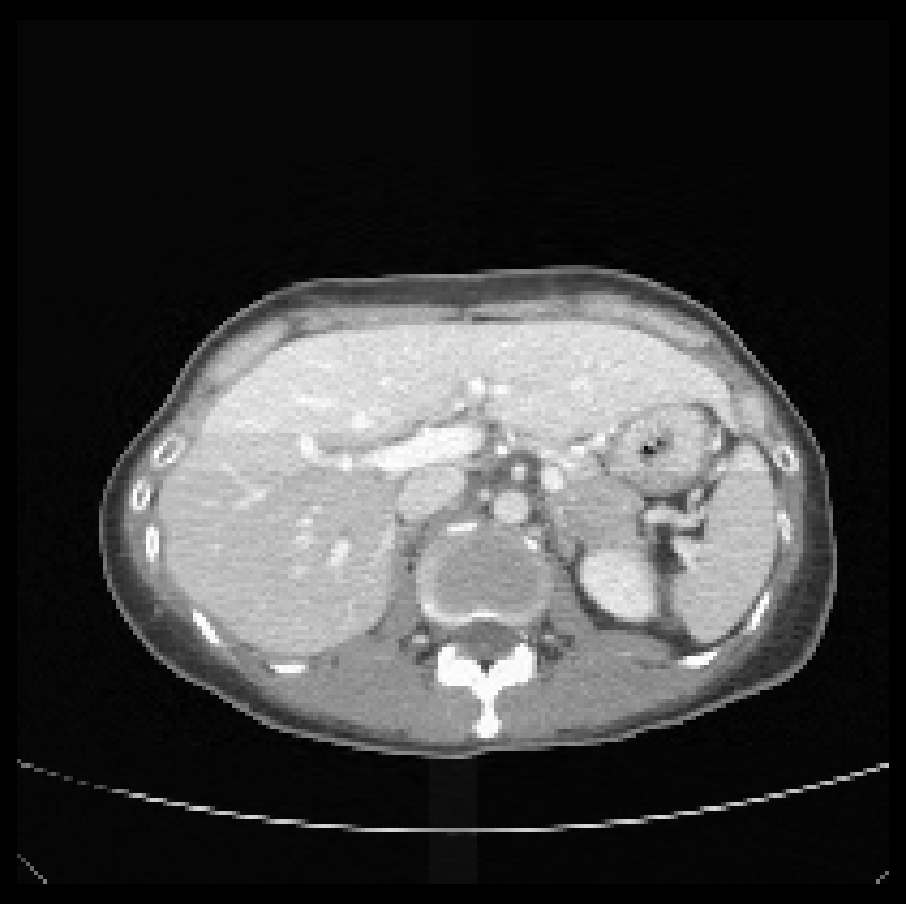}
\end{minipage}
\begin{minipage}{0.25\textwidth}
\centering
\includegraphics[height=2.2cm, width=\linewidth, trim={1cm 4cm 0.5cm 8cm},clip ]{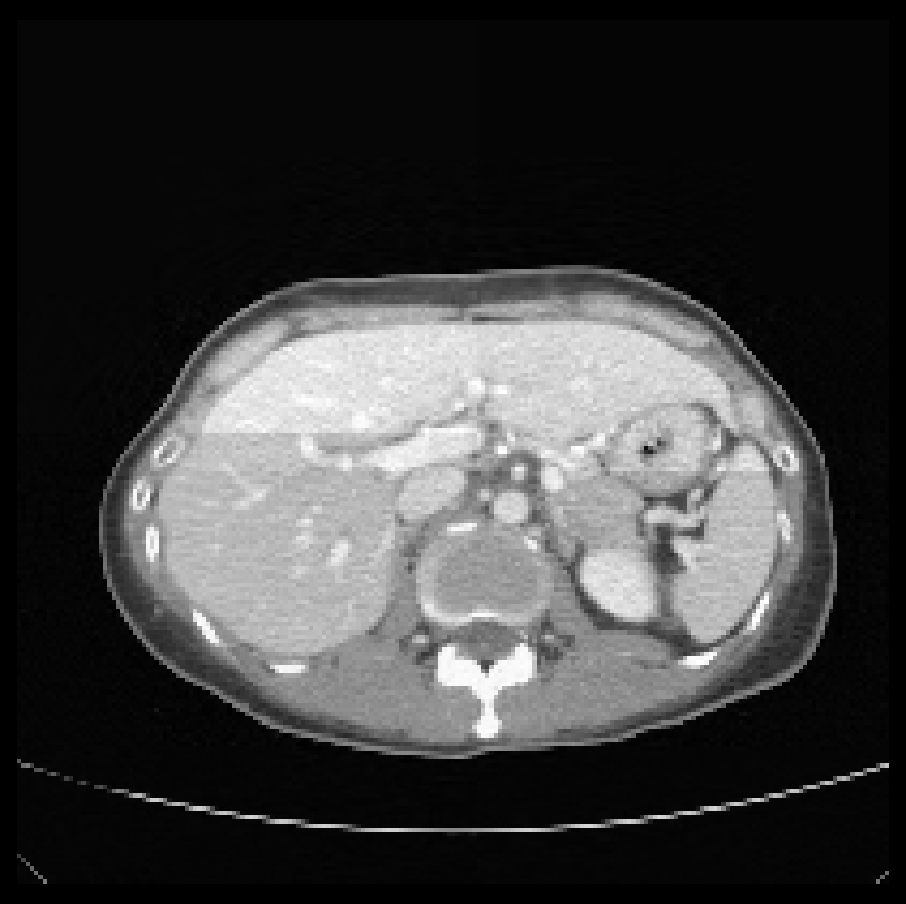}
\end{minipage}
\\
\rotatebox{90}{Synapse}
\begin{minipage}{0.25\textwidth}
\centering
\includegraphics[height=2.2cm, width=\linewidth,,  trim={1cm 4cm 0.5cm 8cm},clip ]{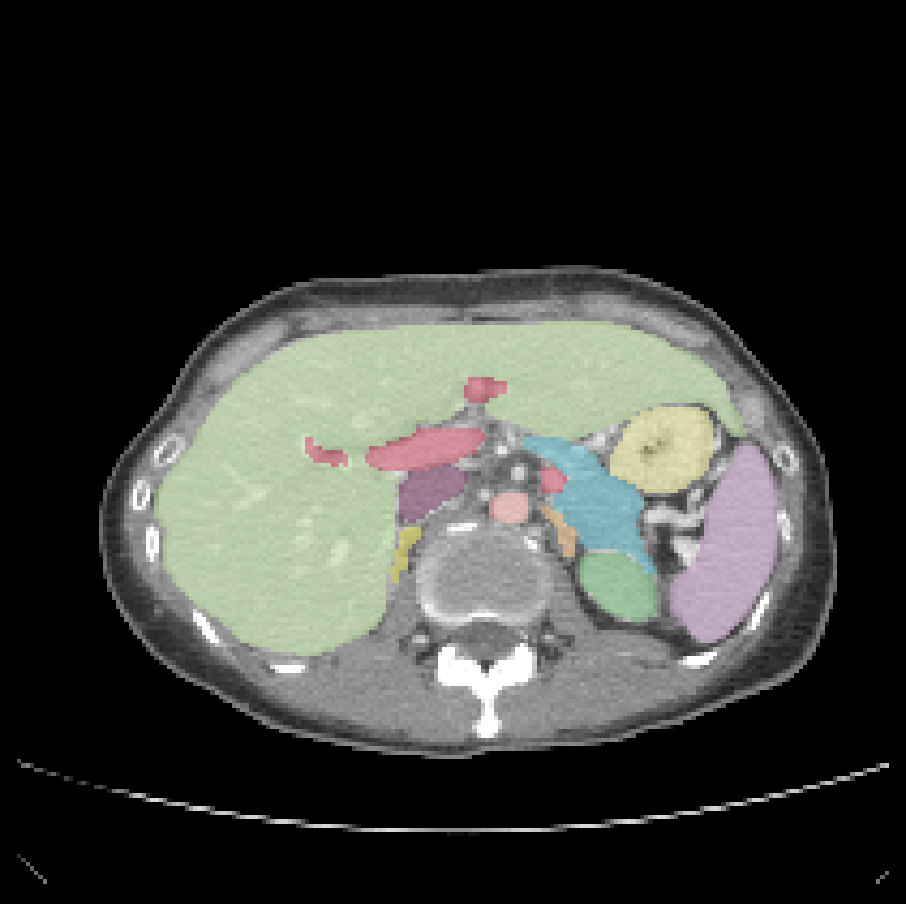}
\end{minipage}
\begin{minipage}{0.25\textwidth}
\centering
\includegraphics[height=2.2cm, width=\linewidth,,  trim={1cm 4cm 0.5cm 8cm},clip ]{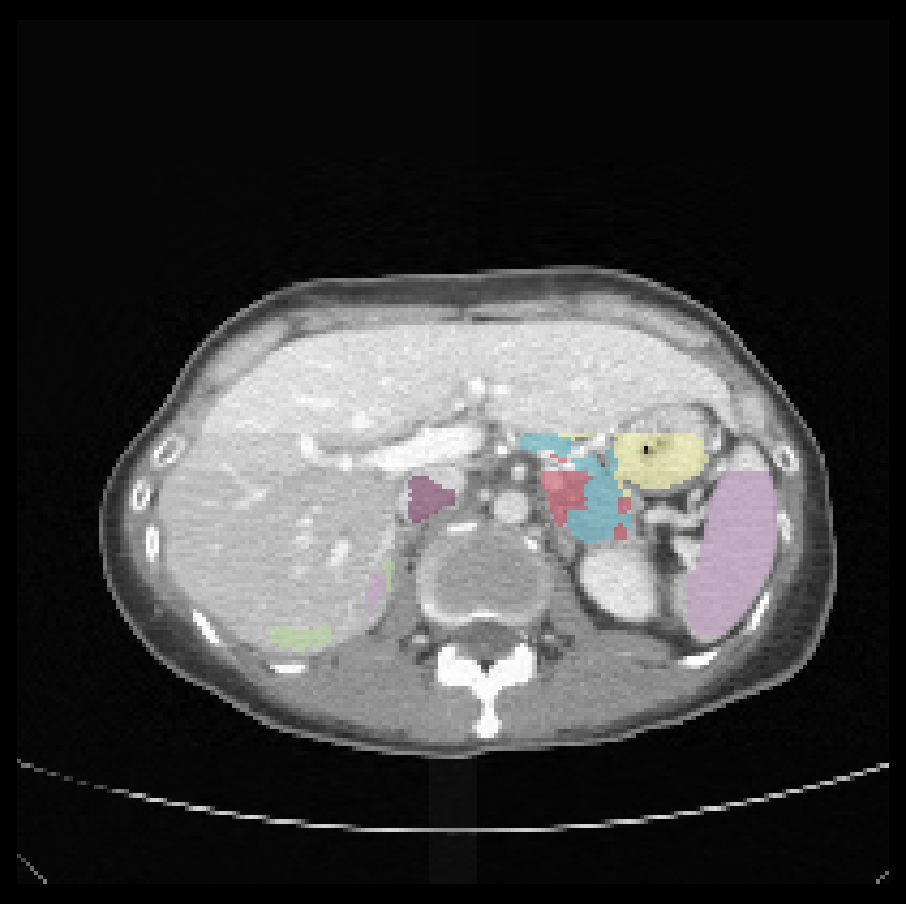}
\end{minipage}
\begin{minipage}{0.25\textwidth}
\centering
\includegraphics[height=2.2cm, width=\linewidth,,  trim={1cm 4cm 0.5cm 8cm},clip ]{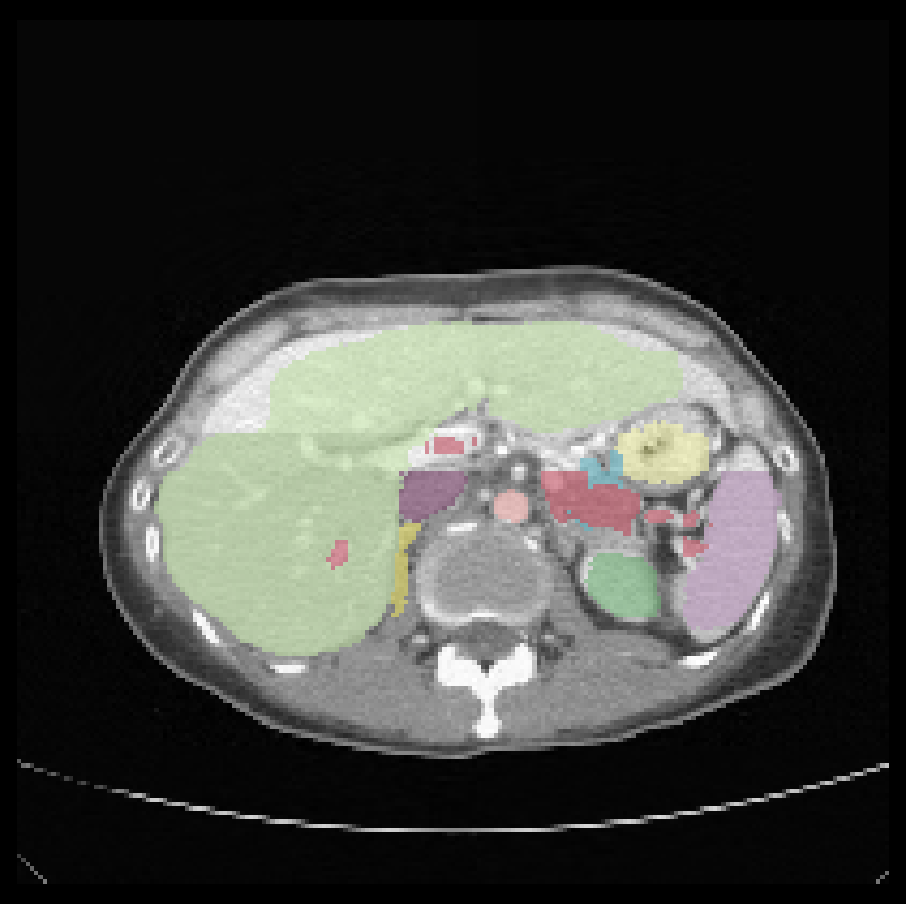}
\end{minipage}
\\
\rotatebox{90}{\phantom{a}ACDC}
\begin{minipage}{0.25\textwidth}
\centering
\includegraphics[height=2.2cm, width=\linewidth,,  trim={1cm 4cm 4cm 2cm},clip ]{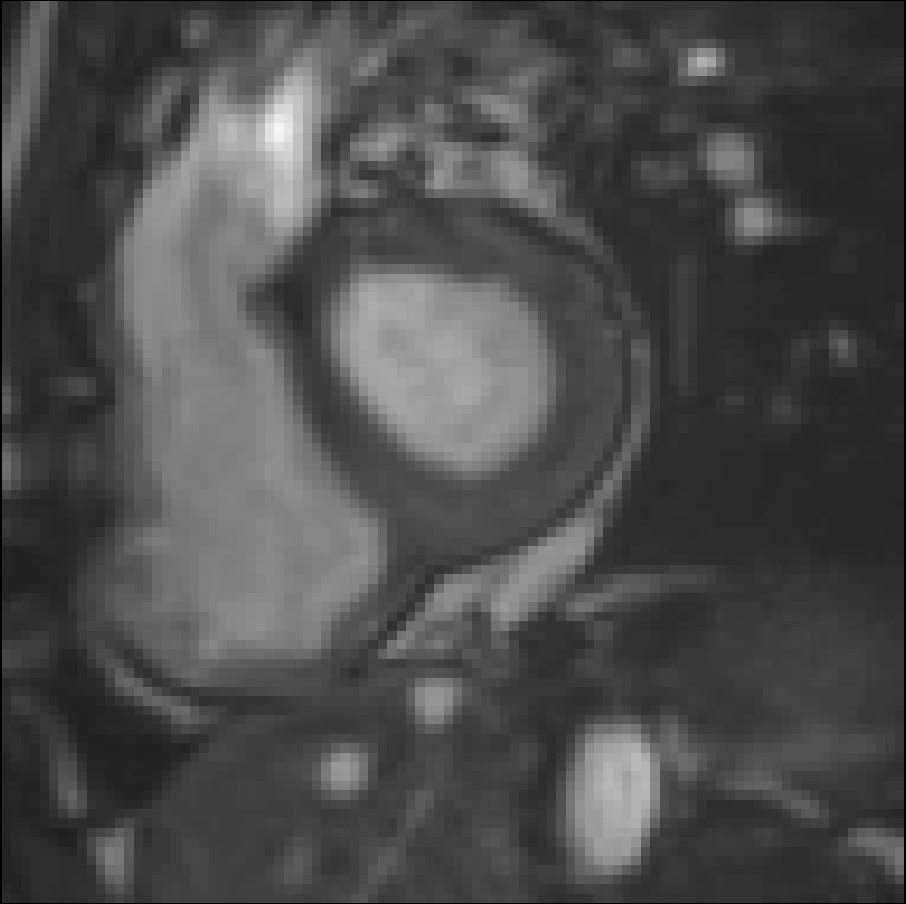}
\end{minipage}
\begin{minipage}{0.25\textwidth}
\centering
\includegraphics[height=2.2cm, width=\linewidth,,  trim={1cm 4cm 4cm 2cm},clip ]{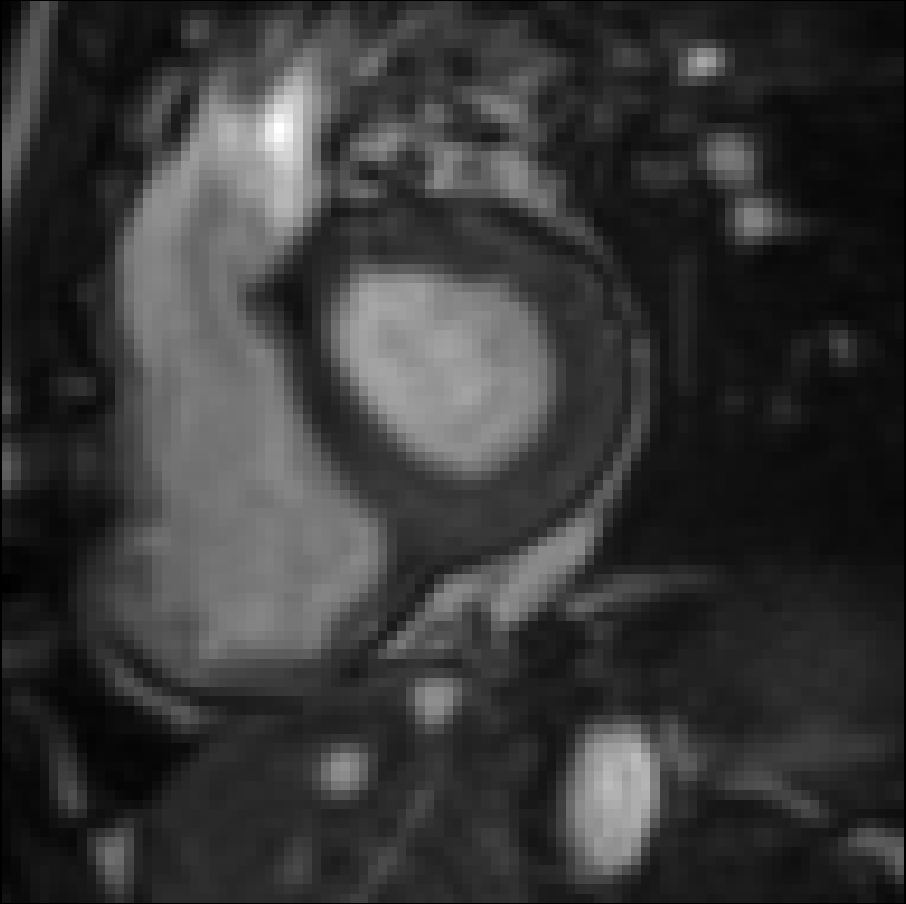}
\end{minipage}
\begin{minipage}{0.25\textwidth}
\centering
\includegraphics[height=2.2cm, width=\linewidth,,  trim={1cm 4cm 4cm 2cm},clip ]{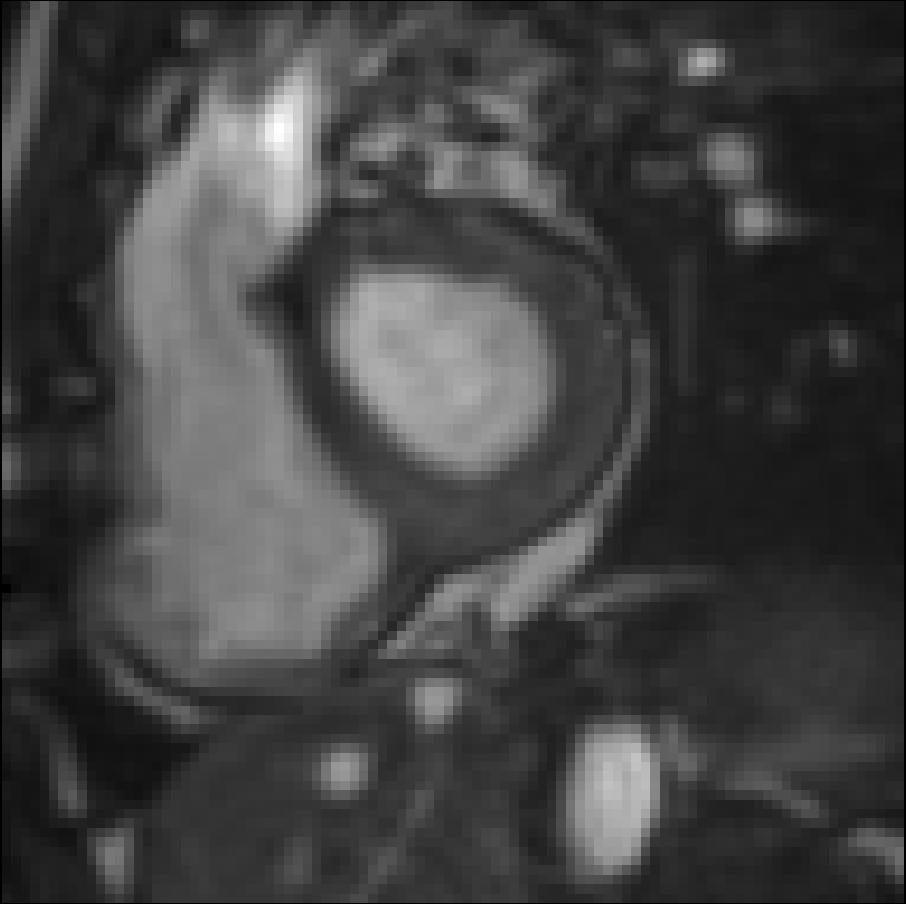}
\end{minipage}
\\
\hspace{0.01em}
\rotatebox{90}{\phantom{abi}ACDC}
\begin{minipage}{0.25\textwidth}
\centering
\includegraphics[height=2.2cm, width=\linewidth, trim={1cm 4cm 4cm 2cm},clip ]{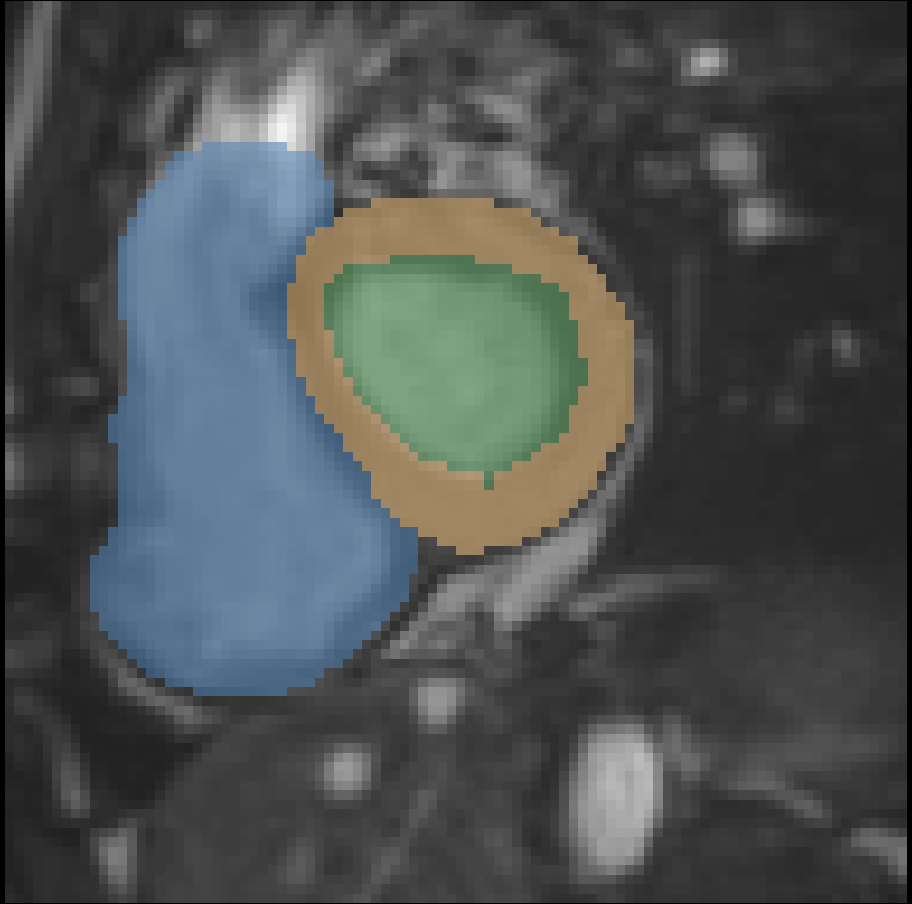}
\footnotesize $\mathrm{Clean}^{\phantom{A}}_{\phantom{A_{a}}}$
\end{minipage}
\begin{minipage}{0.25\textwidth}
\centering
\includegraphics[height=2.2cm, width=\linewidth, trim={1cm 4cm 4cm 2cm},clip ]{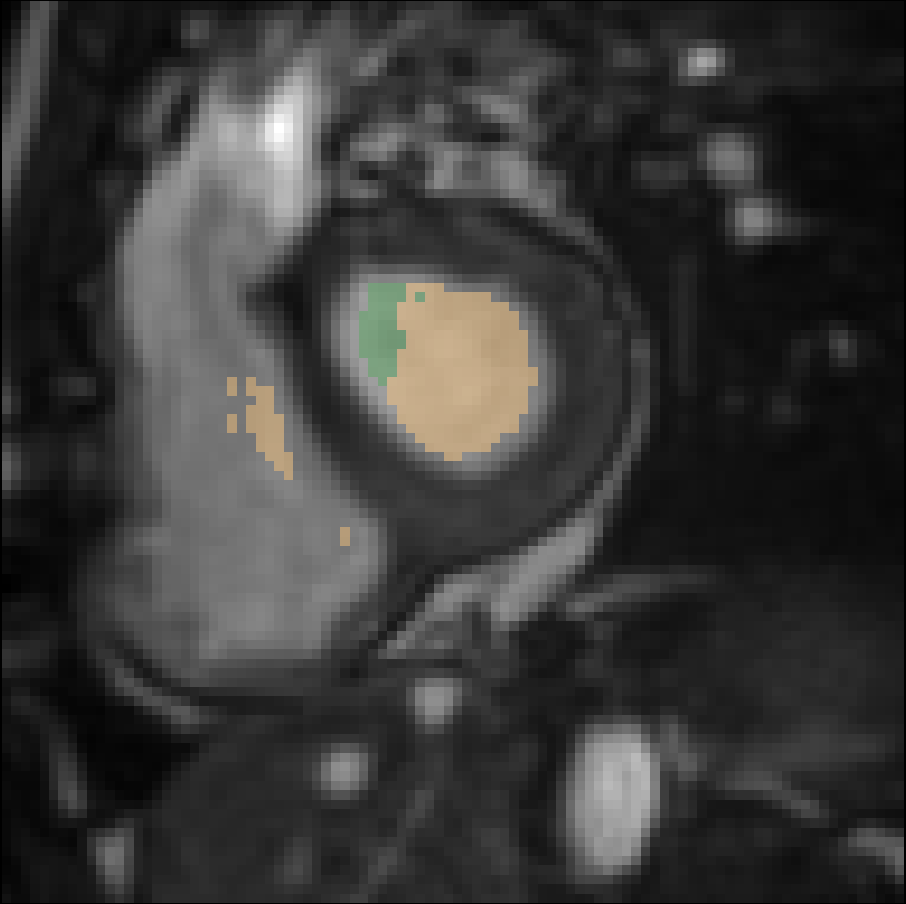}
\footnotesize $\mathcal{M}^{\phantom{A}}_{\phantom{A_{a}}}$
\end{minipage}
\begin{minipage}{0.25\textwidth}
\centering
\includegraphics[height=2.2cm, width=\linewidth,  trim={1cm 4cm 4cm 2cm},clip ]{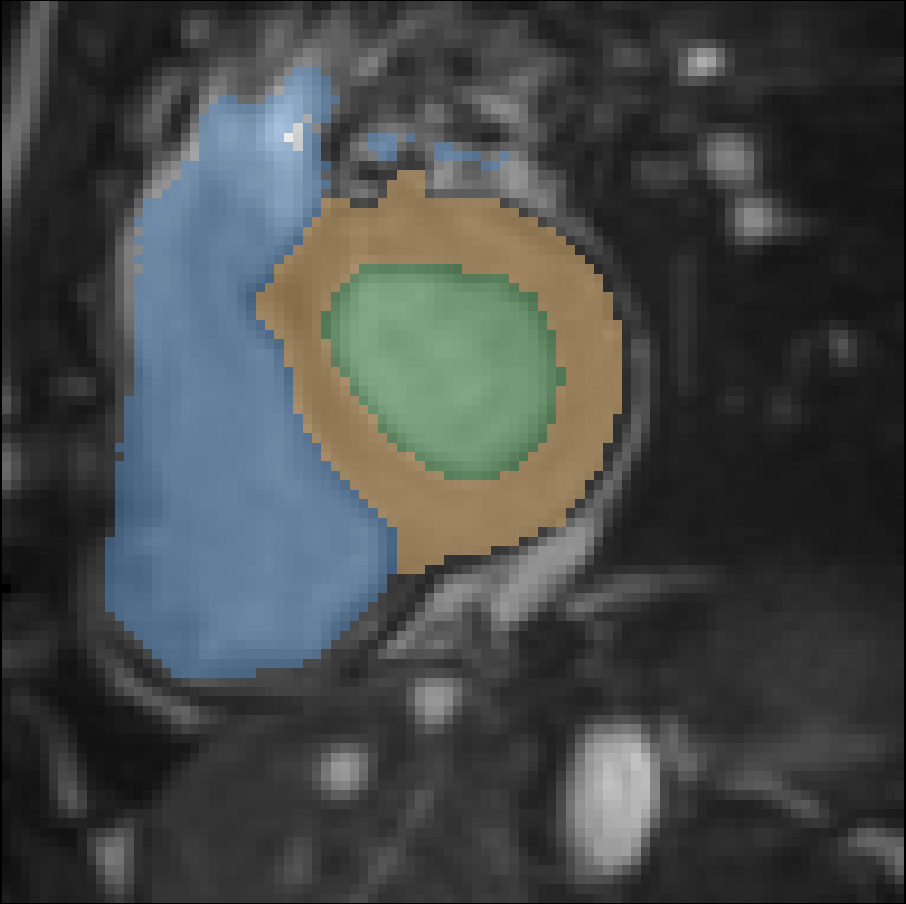}
\footnotesize $\mathcal{M}^{\Scale[0.5]{\text{VAFA-FR}}}_{_{\text{\faShield*}}}$
\end{minipage}
\caption{Qualitative multi-organ segmentation comparison before and after adversarial training of UNETR model under VAFA. First column shows clean image and its ground-truth segmentation labels. Second and third column show images obtained by attacking \textit{vanilla} UNETR model $(\mathcal{M})$ and \textit{robust} UNETR model $(\mathcal{M}^{\Scale[0.5]{\text{VAFA-FR}}}_{_{\text{\faShield*}}})$ which is trained with volumetric frequency adversarial training (VAFT). As compared to vanilla model $(\mathcal{M})$, robust model $(\mathcal{M}^{\Scale[0.5]{\text{VAFA-FR}}}_{_{\text{\faShield*}}})$ has largely recovered the ground truth mask under VAFA.}
\label{fig:before-after-adv-train}
\end{minipage}

\end{figure*}